%% file: paper.tex
\documentclass[sigconf,nonacm=true]{acmart}
\pdfoutput=1

\usepackage{algorithmic}
\usepackage{graphicx}
\usepackage{textcomp}
\usepackage{xspace}
\usepackage{hyperref}
\usepackage{subcaption}
\usepackage{array}
\usepackage{balance}
\usepackage{soul}
\usepackage{multirow}
\usepackage{pifont}
\usepackage{url}
\usepackage{textcase}
\def\BibTeX{{\rm B\kern-.05em{\sc i\kern-.025em b}\kern-.08em
    T\kern-.1667em\lower.7ex\hbox{E}\kern-.125emX}}

\newcommand{\system}{\textit{AutoExecutor}\xspace}

\begin{document}

\title[Predictive Price-Performance Optimization for Serverless Query Processing]{Predictive Price-Performance Optimization for \\Serverless Query Processing}

\author{Rathijit Sen}
\affiliation{%
  \institution{Microsoft}
  \country{}
}
\author{Abhishek Roy}
\affiliation{%
  \institution{Microsoft}
  \country{}
}
\author{Alekh Jindal}
\affiliation{%
  \institution{Microsoft}
  \country{}
\vspace{20pt}
}

\begin{abstract}
\input{abstract}
\end{abstract}

\maketitle

\section{Introduction}
\label{sec:intro}
\input{intro}

\section{Query Resource Allocation}
\label{sec:background}
\input{background}

\section{Price-Perf Model}
\label{sec:pcc-models}
\input{pcc_models}

\subsection{Impact of Total Cores}
\label{sec:total-cores}
\input{total_cores}

\subsection{Training Parameter Model}
\label{sec:ml-model}
\input{ml_model}

\section{\NoCaseChange{\system{}} Integration}
\label{sec:system-integration}
\input{system_integration}

\section{Evaluation}
\label{sec:evaluation}
\input{evaluation}

\section{Conclusions}
\label{sec:conclude}
\input{conclude}

\balance
\bibliographystyle{abbrv}
\bibliography{refs}

\end{document}

%% file: abstract.tex
We present an efficient, parametric modeling framework for predictive resource allocations, focusing on the amount of computational resources, that can optimize for a range of price-performance objectives for data analytics in serverless query processing settings. We discuss and evaluate in depth how our system, \system, can use this framework to automatically select near-optimal executor and core counts for Spark SQL queries running on Azure Synapse. 

Our techniques improve upon Spark's in-built, reactive, dynamic executor allocation capabilities by substantially reducing the total executors allocated and executor occupancy while running queries, thereby freeing up executors that can potentially be used by other concurrent queries or in reducing the overall cluster provisioning needs. In contrast with post-execution analysis tools such as Sparklens, we predict resource allocations for queries before executing them and can also account for changes in input data sizes for predicting the desired allocations.

%% file: intro.tex
Modern clouds have democratized data analytics such that users can easily sign up and get access to the most sophisticated analytics platforms in the cloud.
As a result, most of the complexities in owning and operating these data analytics platforms have been taken care of by the cloud providers, and they offer a way simpler pricing model based on the amount of resources actually used.
The newer serverless query processing models, such as those in AWS Athena~\cite{athena}, Google BigQuery~\cite{bigquery}, Synapse Spark~\cite{synapse-spark}, and Synapse SQL~\cite{synapse-sql}, have further alleviated the need for users to provision any dedicated resources and instead these new serverless approaches automatically allocate resources on a per-query level. 
These trends have also led to data analytics
becoming extremely resource intensive in modern clouds due to the massive amount of data they consume and the complex processing they apply over it.
As a result, it is important for enterprises to manage their total cost of operations (TCO) by reducing their resource consumption and doing more analytics with less resources.

Current approaches for optimizing cloud resources are reactive in nature.
For instance, several efforts have considered recommending SKU or other resource recommendations based on past usage patterns, e.g., SKU recommendations in SQL Server~\cite{sql-sku-rec}.
Others such as Sparklens~\cite{sparklens} analyze the performance of a previously executed Spark query to suggest better resource configurations, e.g., number of Spark executors (worker processes) to use. 
Still other automatic approaches include detecting idle cycles to pause or resume the system as in 
Azure SQL~\cite{azuresql-pause-resume},
auto-scaling the resources either at the cluster or pool level as in Synapse Spark~\cite{synapse-autoscale}, RedShift~\cite{redshift-autoscale}, Snowflake~\cite{sf-autoscale}, etc., or at the query level as in Spark~\cite{spark-dyn-alloc} or even Cosmos~\cite{apollo} based on the availability of spare resources.
Unfortunately, these reactive approaches take several minutes to react~\cite{synapse-autoscale} and many of the optimization opportunities may already be missed. Additionally, reactively adjusting resources during the course of a query execution could even lead to expensive changes in the query plan~\cite{raqo}.
Therefore, apart from the reactive approaches, we also need predictive resource allocation to provide a good starting point in the first place.

\begin{figure}[t]
\centering
\includegraphics[width=0.85\columnwidth]{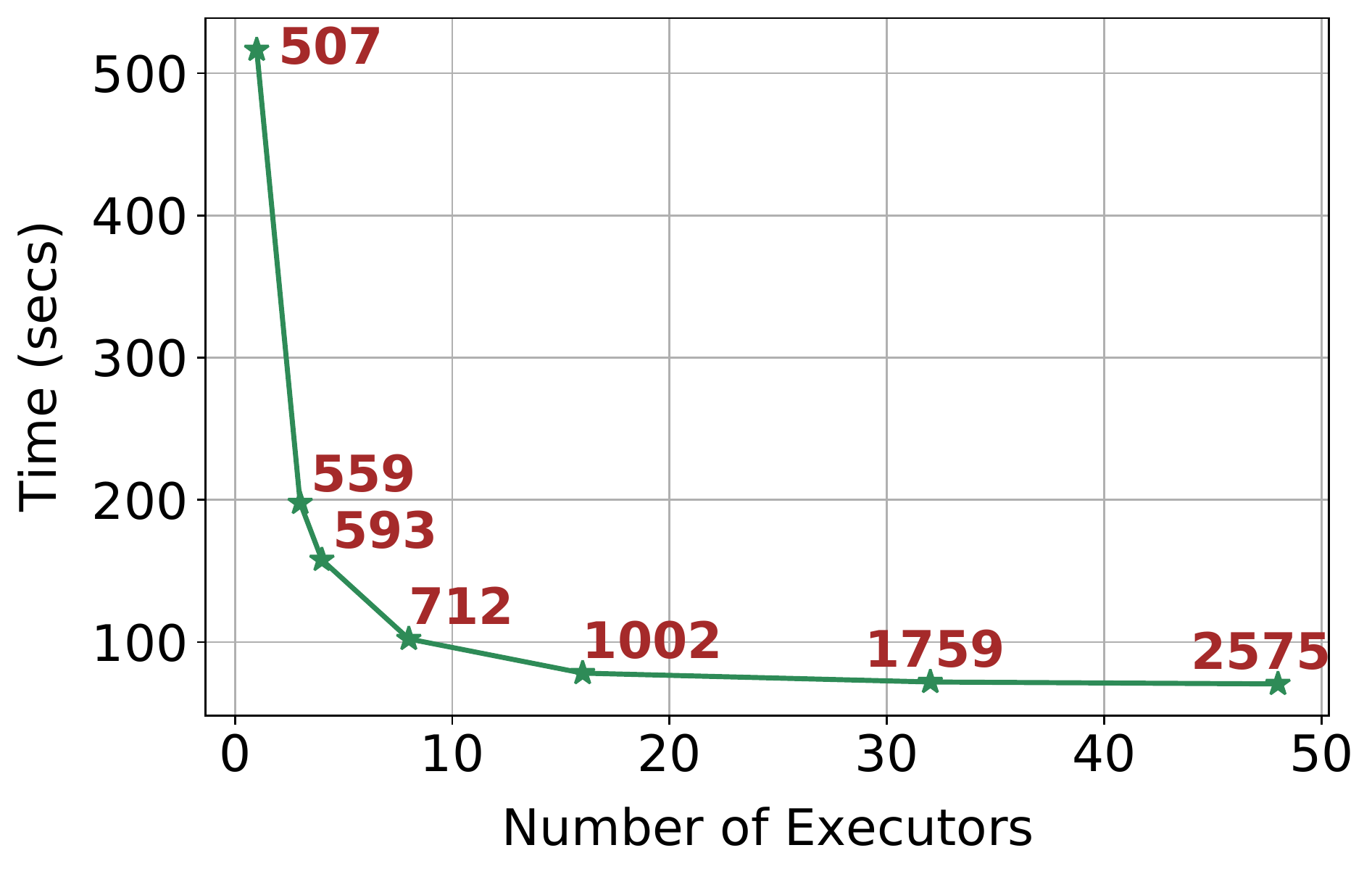}
\caption{Average application run time and area under the executor allocation skyline (data labels) for TPC-DS query94, Scale Factor(SF)=100, when run with different executor counts.}
\label{fig:example-time-auc}
\vspace{-0.3cm}
\end{figure}

Unfortunately, predictively allocating resources to an analytics query is challenging since it is non-trivial to map a query to its resource needs. 
In fact, it is well known that even expert users cannot correctly estimate the resources needed for a given query~\cite{morpheus}.
Furthermore, changing resources for a query impacts its performance, something which is not well understood by users.
To illustrate, Figure~\ref{fig:example-time-auc} shows the performance of TPC-DS query 94 implemented in Spark SQL when using different number of Spark executors, the unit of resources available to queries in Spark.
The green curve shows that the performance improves (lower runtime) initially as more executors are added, however it plateaus later on.
The labels in red show the total executor occupancy (measured in executor-seconds) over the entire query execution.
We see that even when the performance plateaus, the resource consumption continues to increase with more executors.
Thus, we have an interesting price-performance (price-perf for short) optimization at hand.

\begin{figure*}[t]
        \begin{subfigure}[b]{0.33\textwidth}
             \centering
             \includegraphics[width=\textwidth] {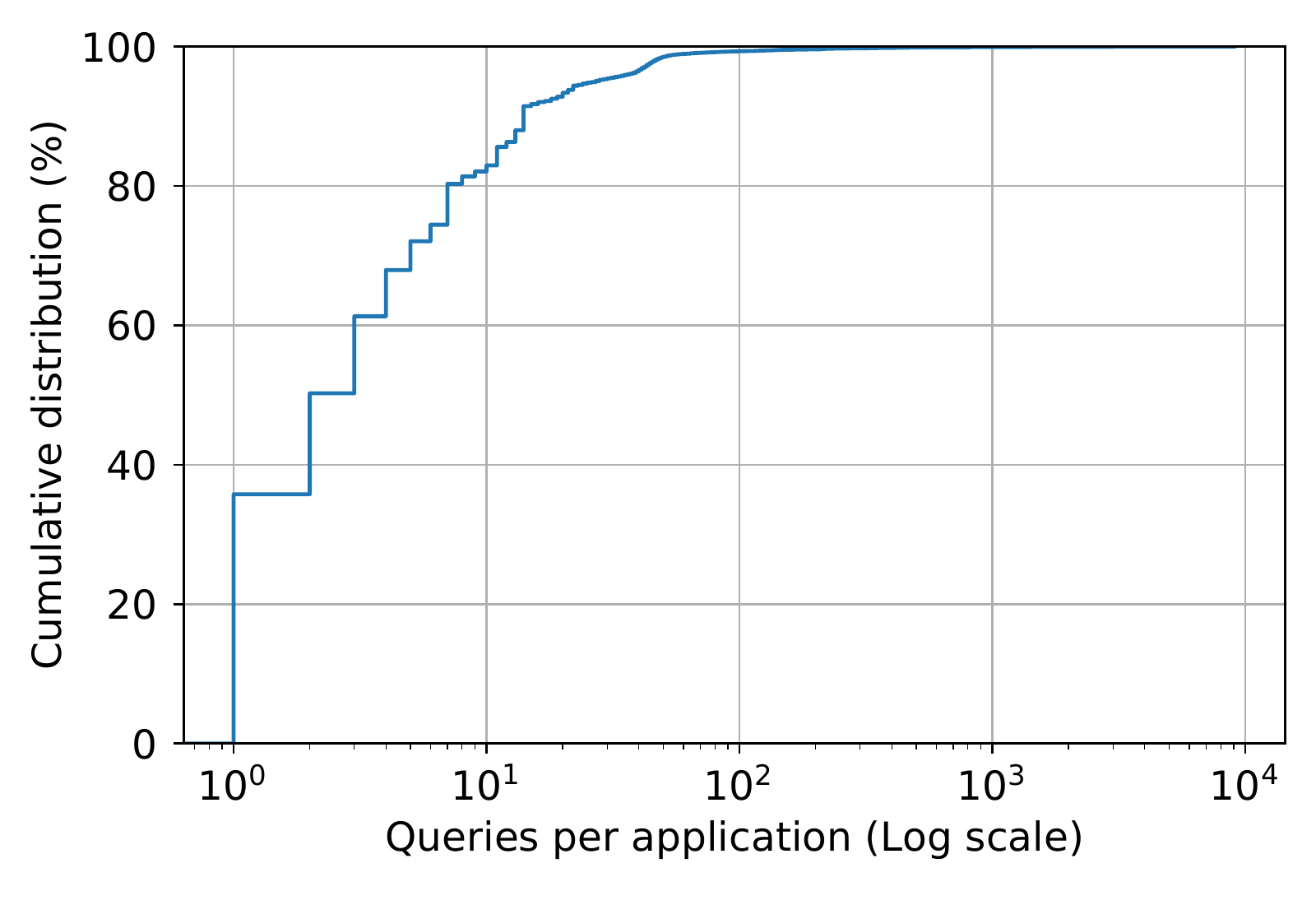}
             \caption{Queries per application}
             \label{fig:queriesPerApp}
         \end{subfigure}\hfill
        \begin{subfigure}[b]{0.33\textwidth}
             \centering
             \includegraphics[width=\textwidth] {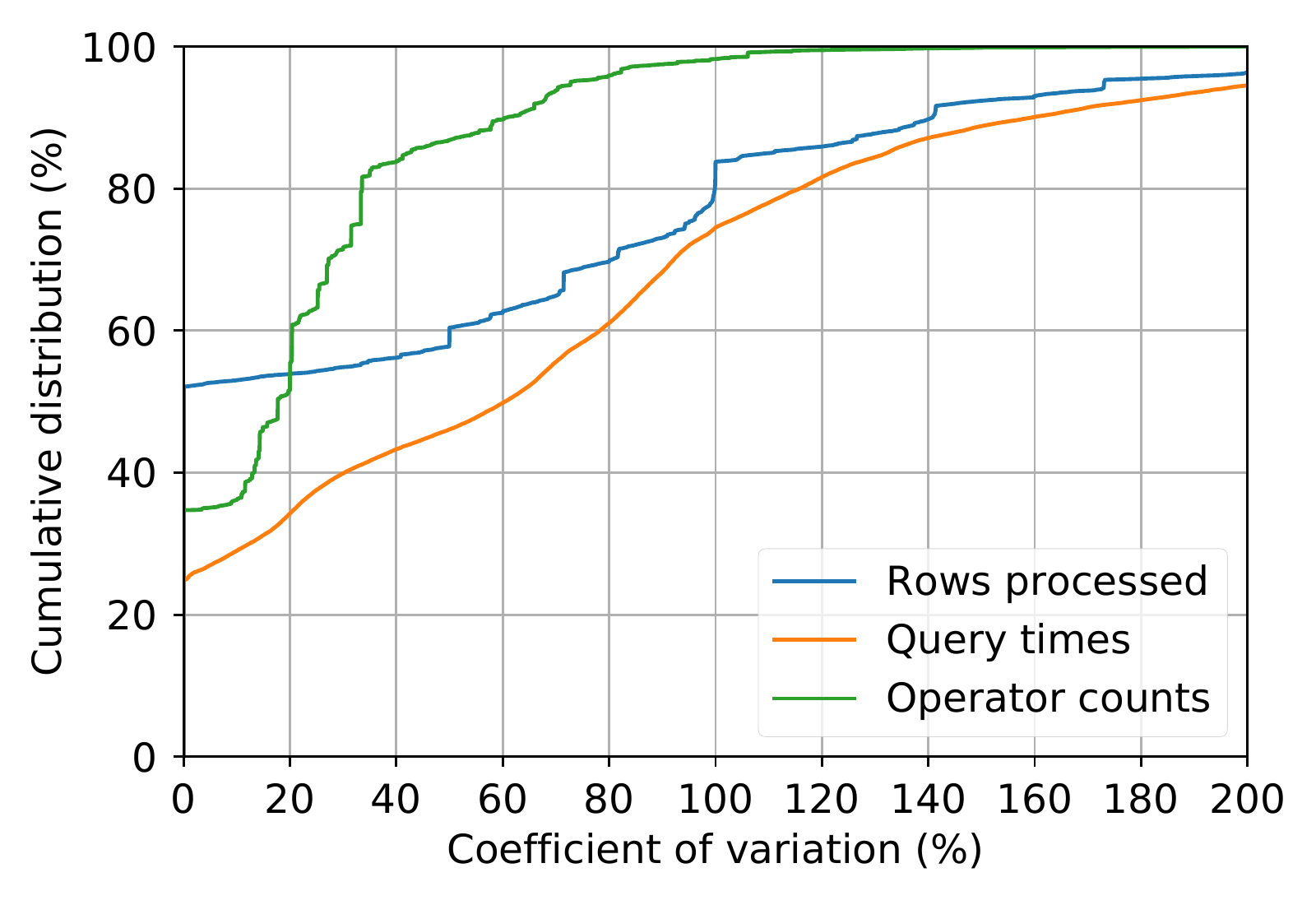}
             \caption{Variation in application queries}
             \label{fig:coeffVar}
         \end{subfigure}\hfill
        \begin{subfigure}[b]{0.33\textwidth}
             \centering
             \includegraphics[width=\textwidth] {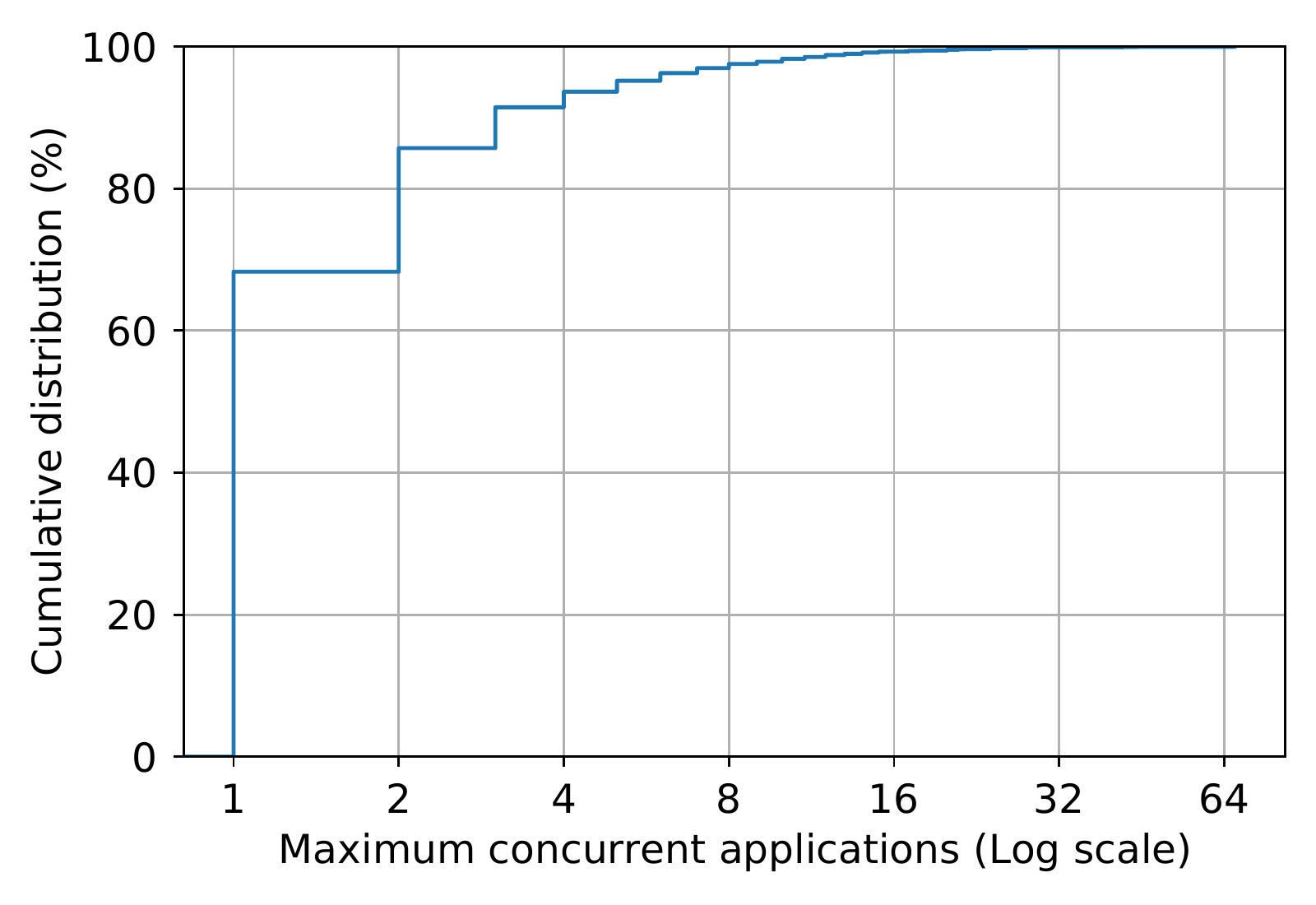}
             \caption{Concurrent applications within clusters}
             \label{fig:concurrentApps}
         \end{subfigure}
    \caption{Insights from production Spark workloads at Microsoft.}
    \label{fig:workload_analysis}
\end{figure*}

In this paper, we study predictive price-perf optimization in serverless query processing setting, i.e., resources are allocated and users are charged at the query level.
We build on top of our prior work on 
the relationship between query performance and resources in Hive and Spark~\cite{raqo},
predictive degree of parallelism in SQL Server~\cite{autodop},
peak~\cite{autotoken}, adaptive~\cite{token-shaper} and optimal~\cite{tasq} resource allocation in SCOPE~\cite{chaiken2008scope} jobs, 
and present an end-to-end framework for predictive price-perf optimization at the query level.
We recently demonstrated
this system design and concept~\cite{autoexecutor-demo}, and in this paper we present more generalized models, detailed architecture, analysis, and results, apart from Spark optimizer extensions that combine both the predictive and the reactive approaches.

Specifically, we make the following key contributions:

\begin{enumerate}

\item We discuss current approaches for query-level resource allocation and use production workload insights from Microsoft to motivate why it is important yet difficult to predictively allocate resources at the query-level. Section~\ref{sec:background}

\item We present a parametric model to analyze the price-perf trade-off and predictively pick the point of operation for a query. We further discuss the impact of the total number of cores, apart from the executor count. We describe how to train the price-perf model parameters using different machine learning models and discuss the feature importance. Section~\ref{sec:pcc-models}

\item We describe, \system, an system that integrates predictive price-perf optimization with the Spark query optimizer, and discuss various aspects including the model format, model lookup, optimizer extensions, and resource deallocation by combining both the predictive and the reactive resource allocation. Section~\ref{sec:system-integration}

\item Finally, we show a detailed set of experiments on predictive price-perf optimizations in Spark SQL using the TPC-DS workload. We discuss prediction accuracy, picking the point of operation, relationship with dynamic allocation, handling workload changes, and the overheads involved. Our results show that on TPC-DS,
\system{} can save executor occupancy by 48\% compared to dynamic allocation's reactive behavior. Section~\ref{sec:evaluation}.

\end{enumerate}

%% file: background.tex
In this section, we discuss various query-level resource allocation approaches. 
We focus our discussion on the Spark query processing engine and consider the executor count, i.e., the number of worker processes, available to each Spark SQL query as a unit of resource since it is a crucial factor in both query performance and cost.

The query processing cost is determined by the resource allocated to execute it. In this work, we focus on the computational resources. Let $n_s$ denote the number of executors allocated to a query at time $s$ during its execution. We are interested in the following two metrics.
\begin{enumerate}
\item The maximum executor allocation, $n = max(n_s)$, that impacts both query performance and total provisioning needs. The query performance is inversely proportional to its total run time, which we denote by $t(n)$.
\item The total executor occupancy, that we denote by $AUC = \int_{s} n_s ds$, and that impacts the total resource reservation. $AUC$ is the Area Under the (skyline) Curve in a timeline plot of $n_s$ vs $s$ over the lifetime of the query execution.
\end{enumerate} 

As Figure~\ref{fig:example-time-auc} shows, both $t(n)$ and $AUC$  are strongly influenced by $n$. In this work,  we do not directly predict $AUC$ or optimize for a specific $AUC$ target, but focus on modeling $t(n)$ as a function of $n$. We refer to this as our Price-Performance Model (PPM). Section~\ref{subsec:eval-auc-savings} shows how our techniques also help to reduce $AUC$.

\begin{figure*}[t]
    \centering
            \begin{subfigure}[b]{0.33\textwidth}
             \centering
             \includegraphics[width=\textwidth] {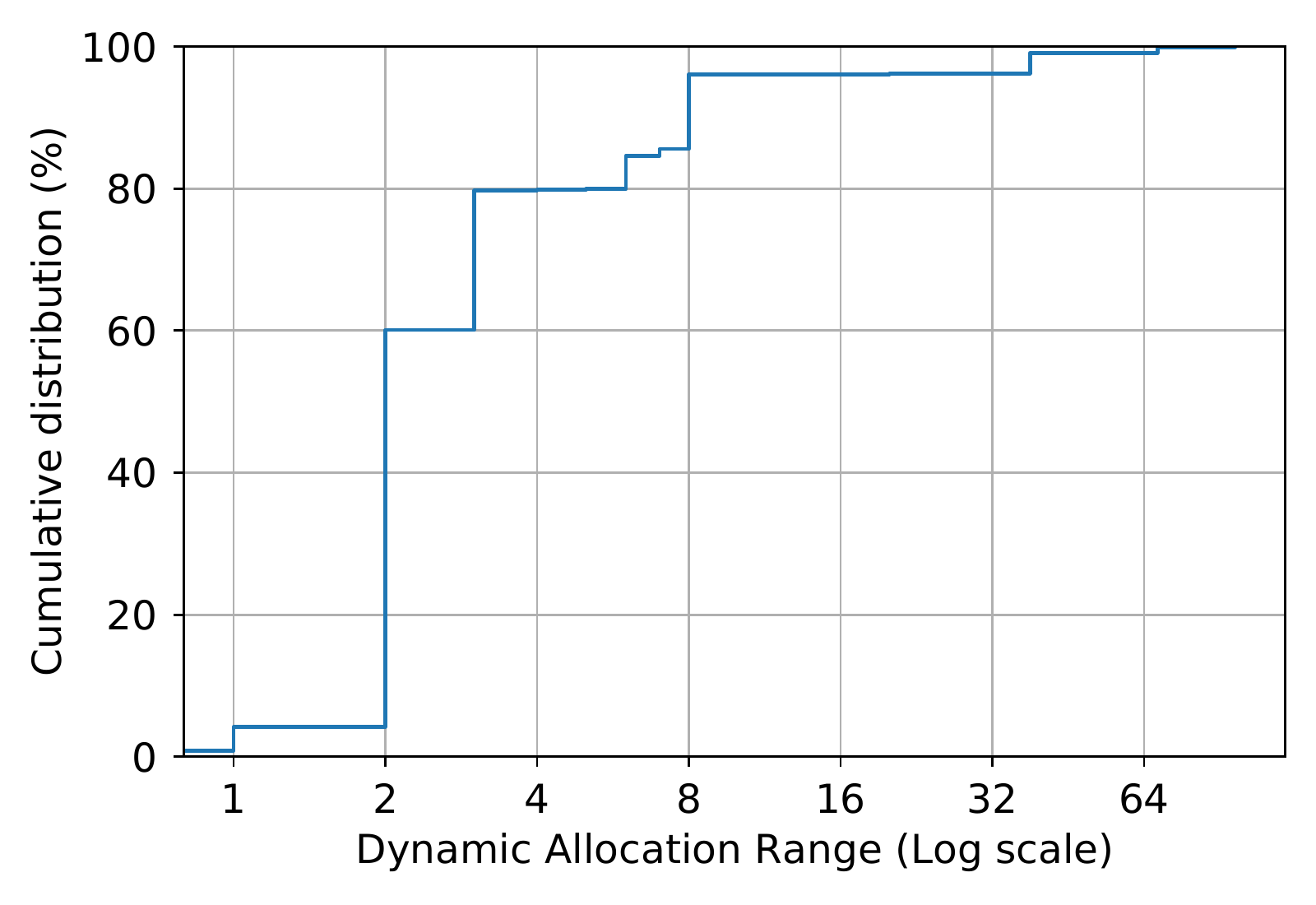}
             \caption{Non-default range for dynamic allocation}
             \label{fig:dynrange}
         \end{subfigure}\hfill
        \begin{subfigure}[b]{0.33\textwidth}
             \centering
             \includegraphics[width=\textwidth] {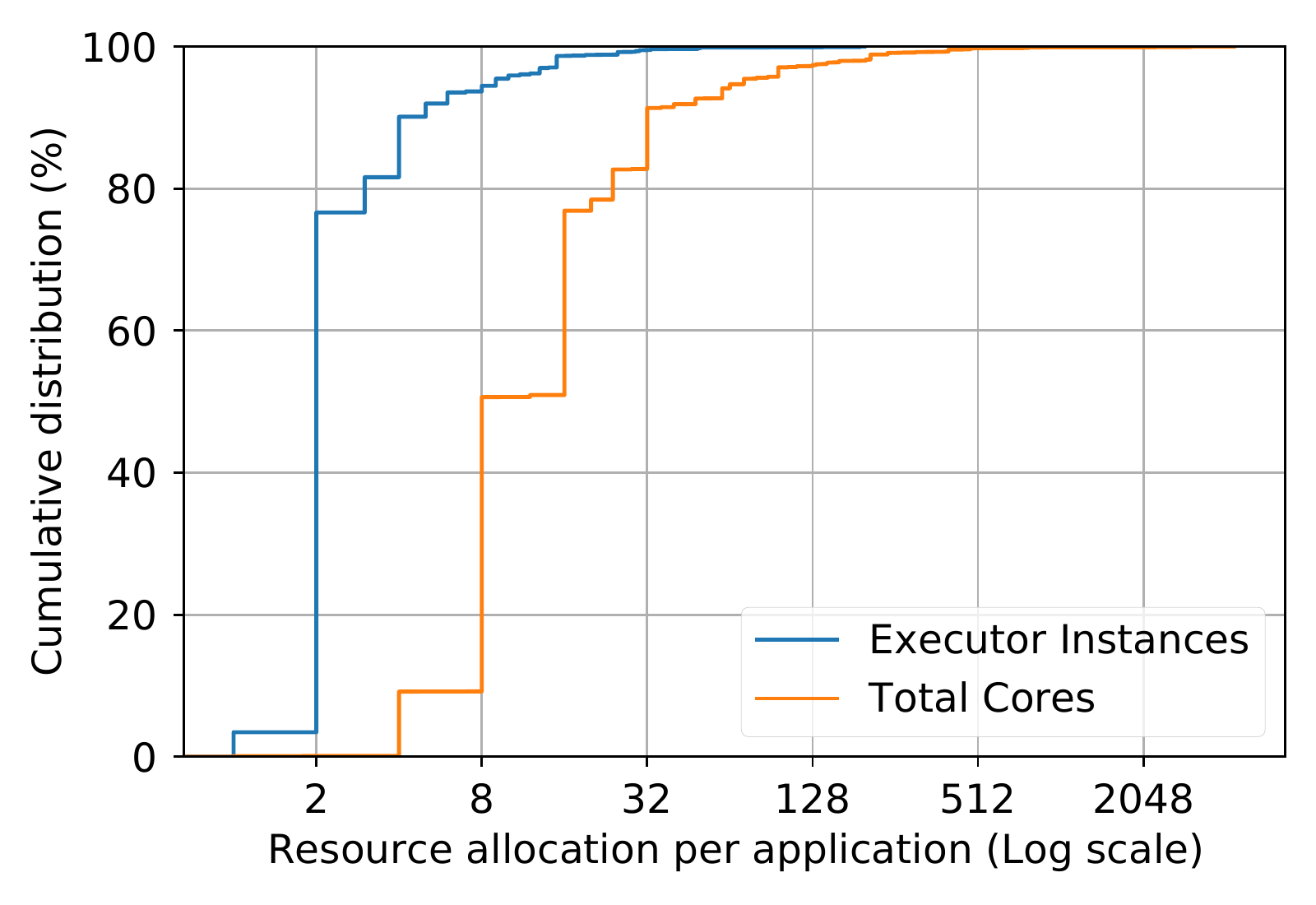}
             \caption{Static resource allocation}
             \label{fig:defaultexec}
         \end{subfigure}\hfill
        \begin{subfigure}[b]{0.34\textwidth}
             \centering
             \includegraphics[width=\textwidth] {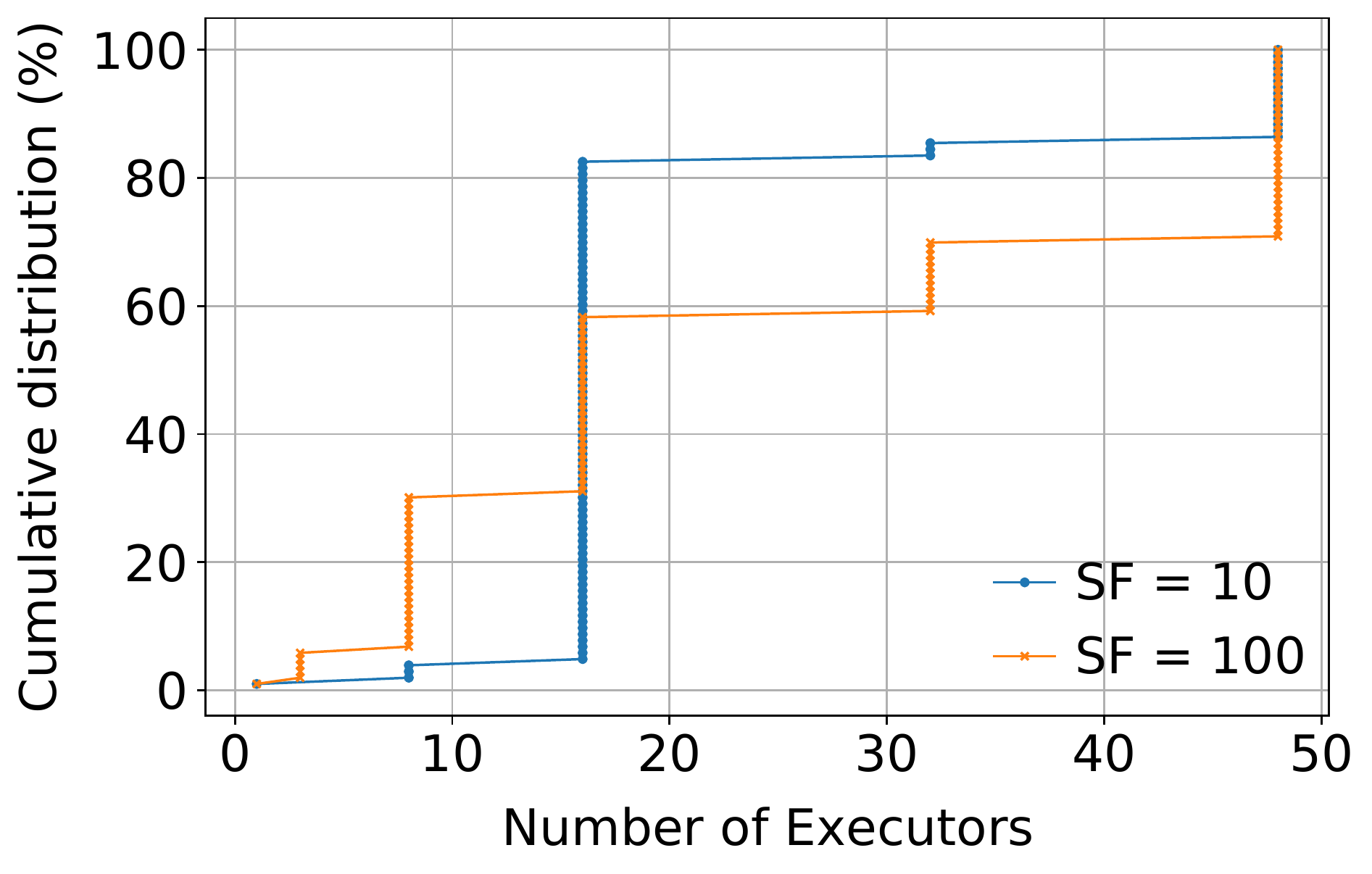}
             \caption{Optimal executors for TPC-DS queries}
             \label{fig:opt-execs}
         \end{subfigure}
\caption{Executor counts in production Spark workloads and optimal executor counts for TPC-DS.}
    \label{fig:default_and_optimal}
\end{figure*}

\subsection{Why Per-query Resource Allocation?}

The first question is why does per-query resource allocation matter. 
We analyze a large subset of daily production Spark workloads at Microsoft consisting of $90,224$ applications and $840,278$ queries across $3,245$ clusters. Figure~\ref{fig:queriesPerApp} shows the distribution of the number of queries per Spark application.
We can see that more than $60\%$ of the applications have more than one query and so all of them need to be considered if we were to allocate resources at the application level.
Furthermore, Figure~\ref{fig:coeffVar} shows that the variation of queries within each of the Spark applications.
We see that queries in half of the applications exhibit a coefficient of variation of $20\%$ or more in their operator counts, $40\%$ or more in the number of input rows processed, and $60\%$ or more in their query execution times. 
As a result, queries within an application are quite different and are expected to have different resource requirements that might be hard to aggregate at the application level.
Finally, Figure~\ref{fig:concurrentApps} shows the number of concurrent Spark applications in a cluster at any given point, and we see that around $70\%$ of the applications do not share compute resources with other applications in the same cluster. The contention for shared resources is further reduced by the disaggregated storage service in Azure Synapse~\cite{adls}. Thus, it is more practical to allocate resources efficiently at per-query level within each Spark application.

\subsection{The Default Behavior}
\label{subsec:default-behavior}

Let us now look at the default behavior when allocating the executor counts for each Spark SQL query.
Our production Spark workloads at Microsoft show that $59\%$ of the Spark applications have dynamic allocation enabled, which is a reactive approach to adjust the executor count based on the pending requests (more on dynamic allocation below). Interestingly, $97\%$ of these applications with dynamic allocation enabled also have the default minimum and maximum executor threshold set, which are $0$ and $2^{31}-1$ respectively.
For the remaining $3\%$  applications users set their own values and Figure~\ref{fig:dynrange} shows the distribution of the executor count range in those applications. 
We can see that almost $60\%$ of these applications have a range of just $2$, while the remaining could have a range growing all the way to $64$ executors. 

The other $41\%$ of the Spark applications that do not have dynamic allocation enabled by default have the default executor count as shown in Figure~\ref{fig:defaultexec}.
We can see that $80\%$ of these applications that do not have dynamic allocation enabled run with a default executor count of $2$. 

Overall, we observe that the default resource settings for Spark SQL queries are far from ideal, with unrealistic minimum and maximum values set for the executor counts.

\subsection{Reactive Approaches}

We now discuss the major reactive approaches to deal with the above default behavior.

First of all, Spark dynamic allocation~\cite{spark-dyn-alloc} reacts to tasks piling up during the course of query execution and allocates exponentially more executors to improve the performance.
While it can indeed help with unexpected number of tasks, it takes some time for the additional executor requests  to be fulfilled, and might require several requests before the required number of executors are finally allocated. Therefore, it runs the risks of allocating too late as well as exponentially overshooting the required count. Users can control the minimum and the maximum executor counts with dynamic allocation, however as we discussed above, the default values are set to $0$ and $2^{31}-1$ respectively.

Sparklens~\cite{sparklens} allows users to analyze executor count usage of a Spark SQL query in a post-hoc manner, i.e., reacting after a query has finished executing for future instances of the same query. Users can construct the skyline of executor count and get recommendations for better executor count allocations.
Still, the approach is limited to improving the resource allocation for the same query and requires analysis for every such query, which may not be possible for large production workloads.
Other approaches such as Tuneful~\cite{fekri:tuneful:kdd:2020}, Reloca~\cite{hu:reloca:infocom:2020}, and Perforator~\cite{perforator} rely on sample runs to generate training data and build models for predicting the resource requirements.

Cloud analytics platforms today provide auto-scaling of the compute nodes, i.e., reactively expand or shrink the total number of nodes available to a customer.
Theoretically, if such auto-scaling is fast enough then it can react to per-query requirements by allocating or de-allocating the overall resources available, i.e., if more tasks get queued up and dynamic allocation cannot find anymore available executors, then auto-scale can kick in and allocate more nodes.
In practice, however, allocating VMs can take several minutes~\cite{synapse-autoscale}, thus taking a while before the auto-scale can react to resource needs at the query-level.

Finally, many analytics platforms also have auto-pause capability to stop charging customers if there are compute resources have been idle for some time~\cite{azuresql-pause-resume,sf-autospause,ec2-predictive-scaling,seagull}.
However, similar to auto-scale, auto-pause could take several minutes to react and comes into effect only when there are no queries by the user or tenant.
Furthermore, auto-pause is even more conservative since the paused nodes will take a few minutes to be resumed making it ineffective for short pause intervals.

\subsection{Prediction Challenges}

Let us now see why predictive resource allocation at the query level is challenging.
Figure~\ref{fig:opt-execs} shows the distribution of the optimal number of executors over different TPC-DS queries for two scale factors. 
We can see that the optimal values vary for different queries and also for the different scale factors, varying from as little as $1$ executor all the way to $48$ executors, and thus indicating that a rich set of features containing both the query and data characteristics are needed to predict the resources at the query level.

%% file: pcc_models.tex
In this section, we present our approach to price-perf model (PPM for short). 
We build on top of prior work on predicting optimal resources for SCOPE jobs at Microsoft~\cite{tasq}, 
and extend it to a more generalized framework for price-perf optimization.
We discuss this framework below.

\subsection{Model Framework}

Our approach to selecting the optimal configuration involves first predicting the PPM, that is, the relationship between resource allocation and execution time and then selecting the optimal configuration according to the price-performance optimization objective. We can use the same predicted PPM to select different configurations that optimize for various objectives without needing to re-predict the PPM separately for each scenario. There are two components to our modeling approach that are defined as follows.
\begin{enumerate}
\item Represent the PPM by a mathematical function with known properties. It is parametrized by scalars whose values depend on the query characteristics. The time $t(n)$ taken by a query with $n$ executors is given by:
\begin{equation}
t(n) = f(n, \{\mathit{scalar}\text{ }\mathit{parameters}\})
\label{eqn:pcc-general}
\end{equation} 
\item Train a parameter model to learn the values of the scalar parameters of $f$ for a given query:
\begin{equation}
g: \text{query characteristics} \mapsto \{\mathit{scalar}\text{ }\mathit{parameters}\}
\label{eqn:parameter-model}
\end{equation} 
This model is used to predict the parameter values of $f$ for a newly-submitted query. Note that the parameter model is scored only once per query, not once per candidate configuration, and $t(n)$ is estimated by evaluating the predicted instantiation of $f$ at different values of $n$. We discuss the parameter model in more detail in Section~\ref{sec:ml-model}.
\end{enumerate}

While selecting candidate functions for $f$, we impose a condition of monotonicity, similar as in the prior work~\cite{tasq}. This condition means that $t(n)$ should be monotonically non-increasing with $n$. This is consistent with user expectations that the run time of a query should not increase with more resources made available to the query. 

In practice, this expectation can be violated in real systems, e.g., due to parallelism overheads on small sizes or skew in input data. However, we still impose the monotonicity constraint on the PPM model due to the following reasons --
(1) It is never cost-efficient to operate in a region where time increases with allocated resources, so accurately modeling that behavior is not needed; 
(2) even in cases when non-monotonic behavior is observed, the overall minimum time often is the same or close to the minimum time in the initial monotonically decreasing region, as we see in Figure~\ref{fig:total-cores-impact-q69}, so optimization objectives relative to minimum times would not be affected;
(3) run-time estimates from Sparklens, that we use to extract parameters for training our models (Section~\ref{sec:ml-model}), are always monotonically non-increasing;
(4) being consistent with user expectations with simple monotonic models helps with explainability of our resource model decisions and with forecasting the future resources provisioning needs.

In this work, we evaluate two candidates for the PPM model function $f$ as follows. 

\textbf{Power Law with saturation}: We leverage the performance characteristic curve from the prior work~\cite{tasq}, which uses a power-law function for $f$, but also extend it by adding a constant term $m$ that reflects a lower bound on the running time of the query. The PPM model is thus formulated as:
\begin{equation}
t(n) = max(b \times n^a, m)
\label{eqn:pcc-pl}
\end{equation} 
This model has three query-specific parameters, $a$, $b$, and $m$, that the ML model will learn and predict. We abbreviate this model by AE\_PL in the rest of this work.

\textbf{Amdahl's Law}: This is inspired by the well-known Amdahl's Law model for computation speedup with increase in resources used for the computations~\cite{amdahl-law}. In this model, the latency is divided into two parts: a fixed component $s$ that is invariant to changes in resource allocation and a scalable component that is inversely proportional to the amount of resources. The PPM model is thus formulated as:
\begin{equation}
t(n) = s + \frac{p}{n}
\label{eqn:pcc-al}
\end{equation} 
This model has two query-specific parameters, $s$ and $p$, that the ML model will learn and predict. We abbreviate this model by AE\_AL in the rest of this work.

\subsection{Comparison with Sparklens}

\begin{figure}[t]
\centering
\includegraphics[width=0.85\columnwidth]{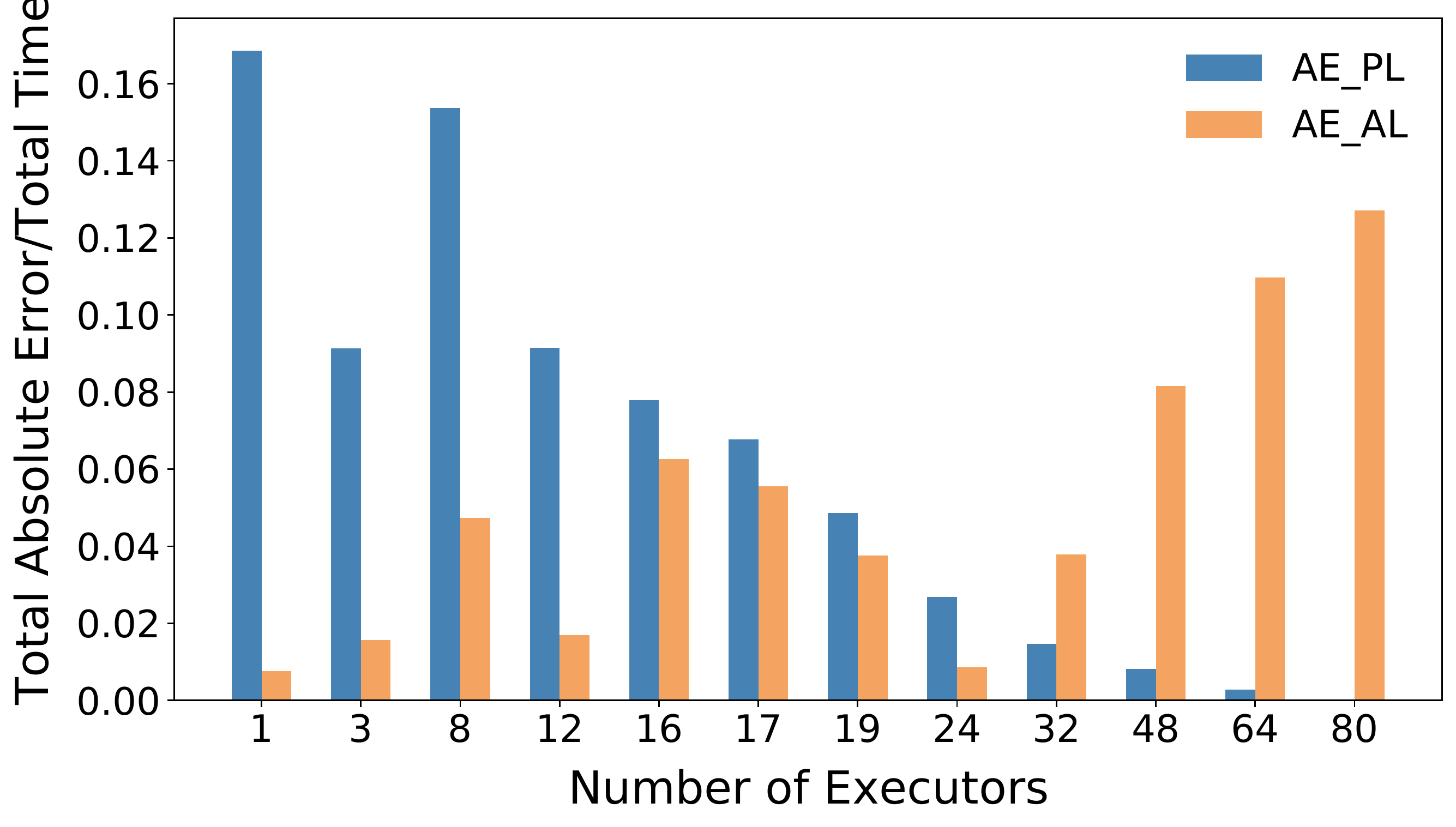}
\caption{PPM model errors for different executor counts for two models, AE\_PL and  AE\_AL, when fit on Sparklens estimates of the PPM over all queries of TPC-DS SF=100.}
\label{fig:model-compare}
\end{figure}

We now compare the accuracy of the above two predictive PPMs with Sparklens, which is a post-hoc reactive model after the queries have finished executing.
While Sparklens simulates the Spark scheduler, it implicitly follows a model of the query run time by first determining the length of the critical path and then distributing the remaining tasks according to available executors. Conceptually, this is similar to the Amdahl's Law approach except that the scaling of time with resources need not be uniform and saturation happens, that is, time estimates stop decreasing beyond some point. 

Figure~\ref{fig:model-compare} shows how well the AE\_PL and AE\_AL models fit Sparklens estimates for all TPC-DS queries at SF=100 for different values of $n$. The Sparklens estimates were obtained from a single run of each query at $n=16$. We see that AE\_AL is a better fit to Sparklens estimates for $n<32$ while AE\_PL is a better fit beyond that. Thus, one can obtain an error of 7\% or less for the full range of $n$ by using the models over different ranges of $n$. Interestingly, we see that although AE\_AL fits Sparklens estimates better at lower values of $n$, it does not translate into better prediction accuracy than AE\_PL when compared to actual query run times in a number of cases, as we will show in Section~\ref{sec:evaluation}.

%% file: total_cores.tex
So far the PPM (Equations~\ref{eqn:pcc-general}--~\ref{eqn:pcc-al}) has considered the number of executors, $n$, as the only input parameter representing the computational resources available to the query. However, the total number of compute cores, $k$, is also determined by the number of cores per executor, $e_c$, since $k = n \times e_c$. To allow for different values of $e_c$, one approach could be to modify the PPM to consider it as an additional parameter, but this increases the model complexity. Instead, we can directly use $k$ in the PPM as we discuss below.

\begin{table}
\centering
\caption{Configurations used for observing impact of total cores on query run time.}
\begin{tabular}{|c|c|c|}\hline
\textbf{Cores/Executor} ($\mathbf{e_c}$) & \textbf{Executors} ($\mathbf{n}$) & \textbf{Total Cores} ($\mathbf{k}$)\\\hline\hline
\multirow{2}{*}{2}& 3& 6\\\cline{2-3}
& 16& 32\\\hline
\multirow{6}{*}{4}& 1& 4\\\cline{2-3}
& 3& 12\\\cline{2-3}
& 4& 16\\\cline{2-3}
& 8& 32\\\cline{2-3}
& 16& 64\\\cline{2-3}
& 32& 128\\\cline{2-3}
& 48& 192\\\hline
\multirow{2}{*}{6}& 3& 18\\\cline{2-3}
& 16& 96\\\hline
\multirow{2}{*}{8}& 3& 24\\\cline{2-3}
& 16& 128\\\hline
\end{tabular}
\label{tab:core-configs}
\end{table}
\begin{figure*}[ht]
\centering
\subfloat[Query 94]{\includegraphics[width=0.33\textwidth]{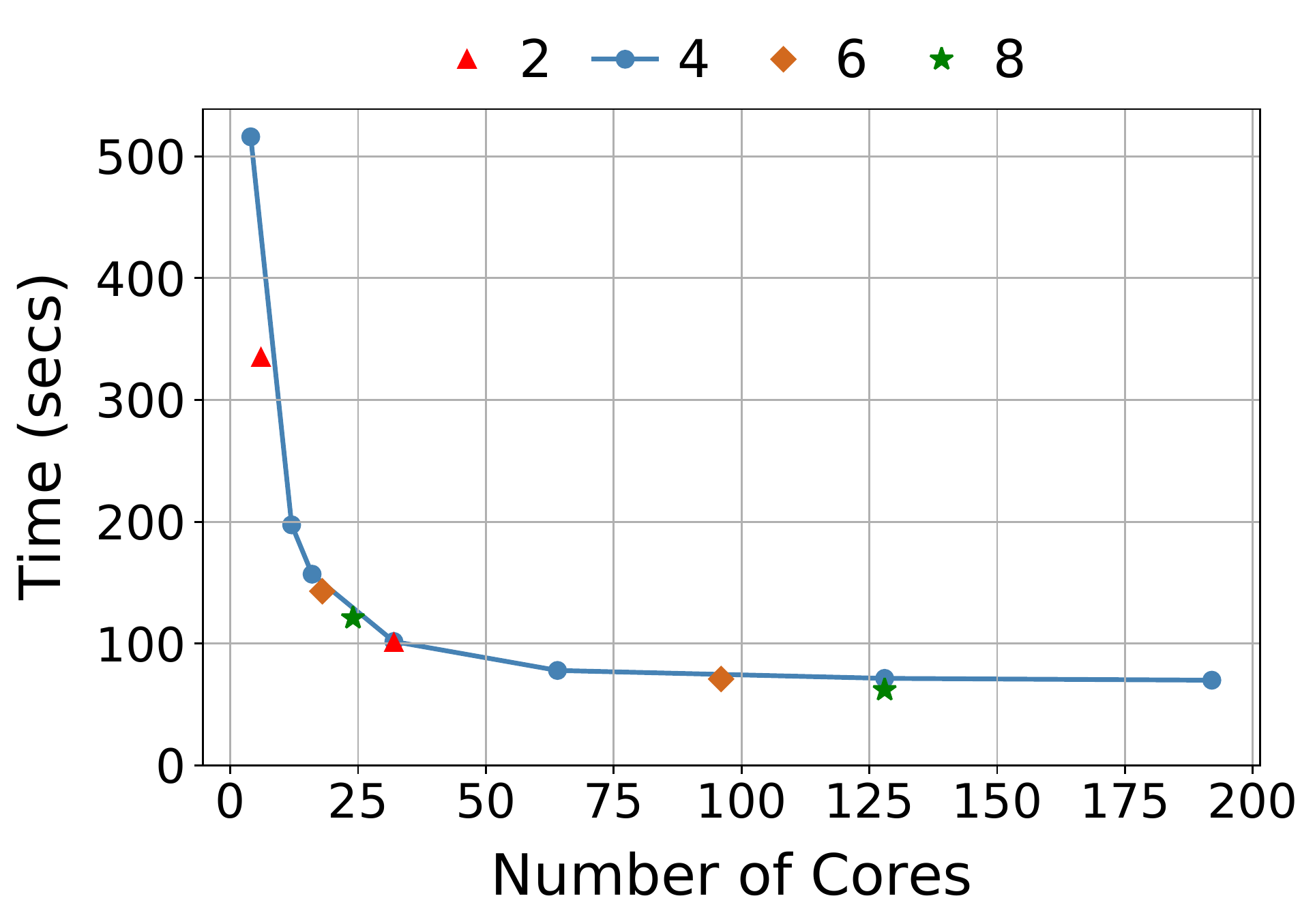}\label{fig:total-cores-impact-q94}}\hfill
\subfloat[Query 69]{\includegraphics[width=0.33\textwidth]{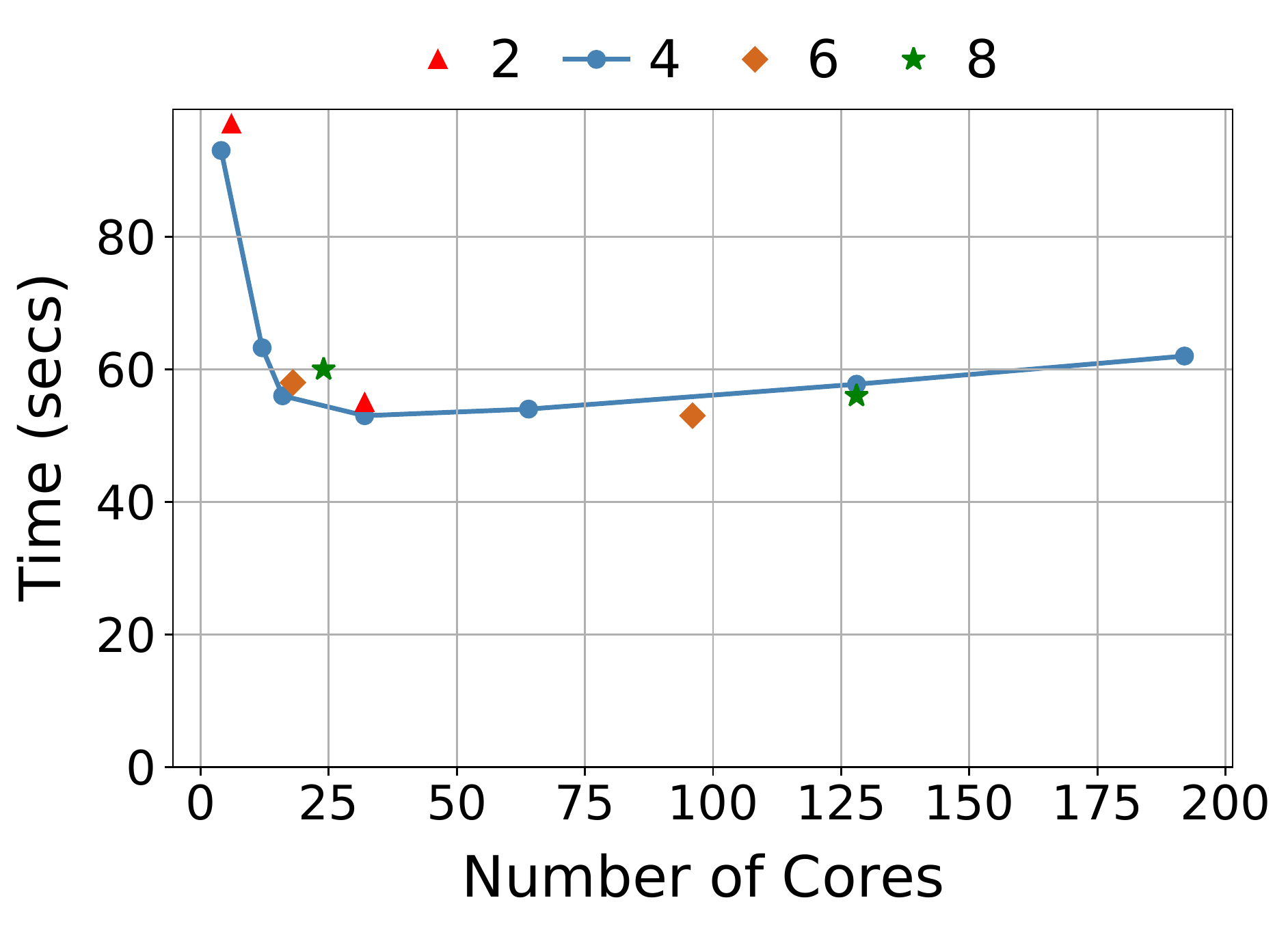}\label{fig:total-cores-impact-q69}}
\subfloat[Error Distribution]{\includegraphics[width=0.33\textwidth]{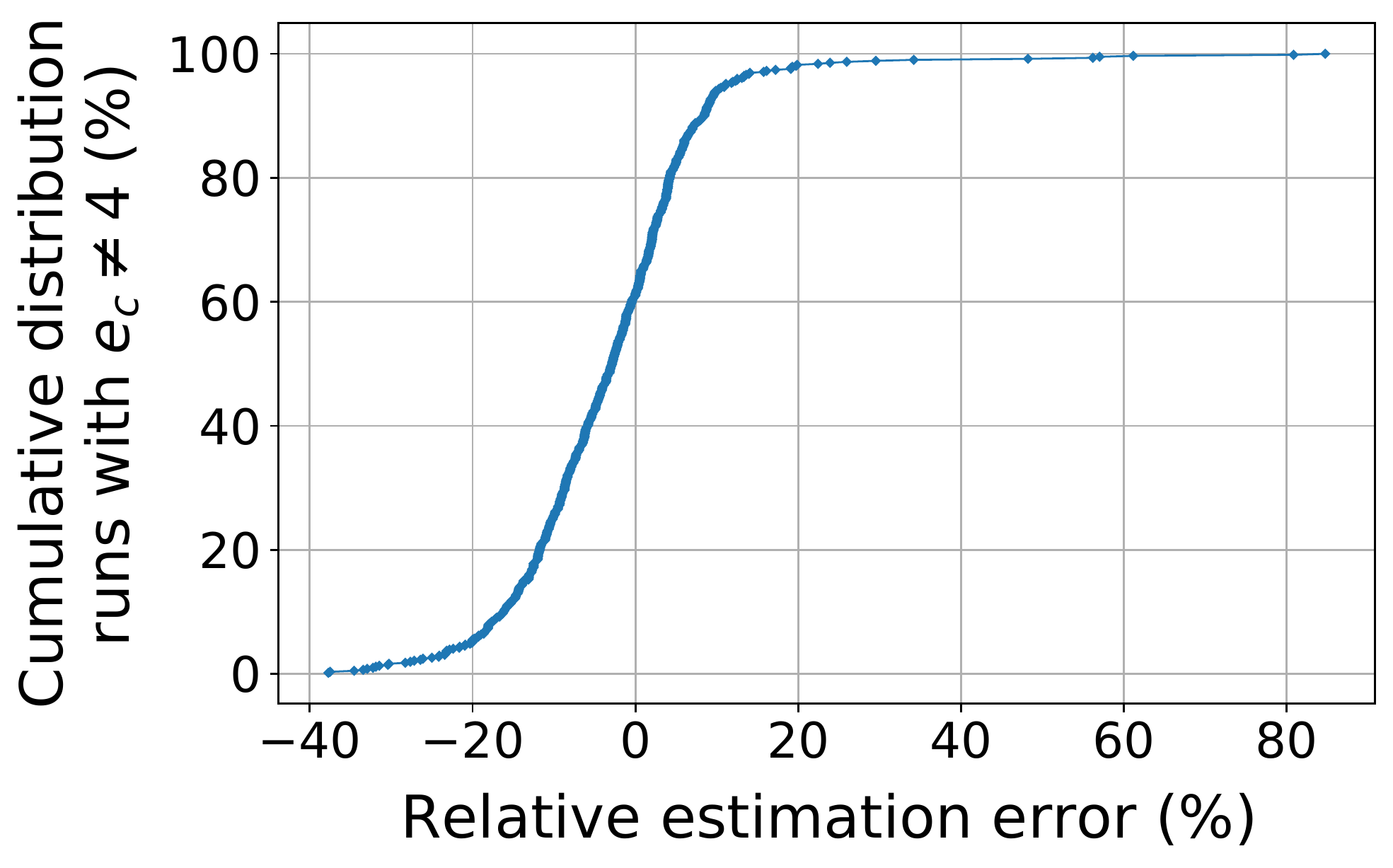}\label{fig:cores-rel-err}}
\caption{(a) and (b) show run times for two example TPC-DS queries, SF=100, with different number of executors, $n$, and cores per executor, $e_c$ (also see Table~\ref{tab:core-configs}). Configurations with the same $e_c$ value belong to the same series. The legend entries show the different series (different $e_c$ values). (c) shows distributions of the relative errors in run time estimation for configs with $e_c\neq 4$, with estimation using linear interpolation from the run times for $e_c=4$, for all queries of TPC-DS SF=100.}
\label{fig:total-cores-impact}
\end{figure*}

To explore this idea, we evaluate several configurations with different values of $n$ and $e_c$ as shown in Table~\ref{tab:core-configs}. Figure~\ref{fig:total-cores-impact} shows actual run times of two example queries from TPC-DS, SF=100, for these configurations. We group configurations into series by the $e_c$ values while the x-axes show the values for $k$. Figure~\ref{fig:total-cores-impact-q94} shows that all the points line up almost on top of the series for $e_c=4$ suggesting that the value of $k$ matters more to overall performance rather than how it is factorized into $n$ and $e_c$. Although the points in Figure~\ref{fig:total-cores-impact-q69} are not fully aligned, the distances from the $e_c=4$ trend line are small.

Figure~\ref{fig:cores-rel-err} shows the distributions of the relative distances from the linearly-interpolated $e_c=4$ values for configurations with $e_c\neq 4$, for the same value of $k$. According to Table~\ref{tab:core-configs}, there are six configurations with $e_c\neq 4$ for each query, and Figure~\ref{fig:cores-rel-err} includes $6\times 103$ points over all the queries. The relative estimation error is calculated as $1-\left(\frac{t_n,e_c\neq 4}{t_n, e_c=4}\right)$ and shown as a percentage. Most estimation errors are small. The relative errors have an average of 8.8\%, with 68.4\% of the points lying in the interval [-10\%,10\%] and 92.9\% of the points are within the interval [-20\%,20\%]. Overall, using $k$ for the PPM instead of $n$ and $e_c$ separately gives good accuracy while reducing model complexity.

Once the optimal value of $k$ is determined for a query, there may be several choices for factorizing it into $n$ and $e_c$ values. Choosing smaller values of $e_c$ offers more granularity in cost-performance tradeoffs due to the larger range of possible values for $n$. Since currently executors cannot span multiple nodes, we want to choose executor sizes that minimizes resource wastage (stranded resources on a node). For example, assume that each node has $C$ cores, $M$ amount of memory, and each executor will get $e_m$ memory. One may solve the following optimization problem.
\begin{eqnarray*}
minimize & C\bmod e_c\\
such \text{ } that& \text{ } e_m \times \lfloor \frac{C}{e_c} \rfloor \leq M\\
and& e_c \times \lfloor \frac{C}{e_c} \rfloor = k
\end{eqnarray*}

Additional considerations can constrain the factorizing strategy, such as reducing garbage collection overheads with very large $e_c$ and avoiding difficulties in determining the optimal amount of overhead memory with very small $e_c$ values~\cite{caffey:exec-config-spark:medium:2020}.

%% file: ml_model.tex
The goal of the parameter model is to learn a function $g:$ query characteristics $\mapsto$ $\{a, b, m\}$ for the AE\_PL model or $g':$ query characteristics $\mapsto$ $\{s, p\}$ for the AE\_AL model, depending on the choice of the PPM. Section~\ref{sec:system-integration} discusses how we extract the query characteristics. For model training, we additionally need the PPM parameters as targets (labels). On the other hand, model scoring will predict the PPM parameter values, and thereby also the PPM. We use an off-the-shelf implementation~\cite{scikit-learn, sklearn-RF} of Random Forest regression models for the parameter model.

For training the parameter model, the PPM parameters are obtained by fitting the PPM to run times of each query for different configurations, which in our case is the number of executors, $n$, or number of cores, $k$. The run times may be from actual query runs or estimates of run times from simulators or other tools. Getting actual run times for different configurations, along with multiple runs to account for run-to-run variance, can be time consuming (also see Section~\ref{subsec:eval-setup}). 
Moreover, rerunning queries with different configurations may not be easy for production workloads. Instead, our approach is to run the training queries once (at $n=16$) and use estimates generated by Sparklens with a post-execution analysis on the logs. 
Sparklens can also generate estimates from production workloads. 
Note that we are using simulation to augment the training data for speed and convenience in production environments, but our models can also be trained with actual run time data for all configurations in case they are available.

Once the training data is available, the PPM parameters for each query can be obtained as follows. For the AE\_PL model, $m$ is the minimum run time over all the different configurations. The power-law portion of the PPM can be transformed into logarithmic space:
\begin{equation}
log(t(n)) = log(b) + n\times log(a)
\end{equation}
$log(a)$ and $log(b)$ can then be determined by fitting a linear regression model to $log(t(n))$ as a function of $n$. For this, we consider only the non-saturating region for $t(n)$, that is, over the region $n\in[1,n_m]$ where $t(n_m)=m$ for all $n>n_m$.

For the AE\_AL model, $s$ and $p$ can be determined by fitting a linear regression model to $t(n)$ as a function of $\frac{1}{n}$.

\begin{table}[t]
\centering
\caption{Feature list for parameter model}
\label{tab:model-features}
\begin{tabular}{|>{\centering\arraybackslash}m{0.38\columnwidth}|>{\centering\arraybackslash}m{0.52\columnwidth}|}\hline
\textbf{Feature}& \textbf{Description}\\\hline
\# Aggregate, Project, Join, Filter, Sort, Union, etc.
& Count of each type of operator in the query plan (14 operators for TPC-DS)\\\hline
$\sum$ all operators& Total number of operators in the query plan\\\hline
Max Depth & Maximum depth of query plan\\\hline
\# Input sources & Number of input data sources used by the query\\\hline
$\sum$ Input bytes & Estimated total number of bytes of input data used by the query\\\hline
$\sum$ Rows processed & Estimated total number of rows processed by all operators in the query\\\hline
\end{tabular}
\end{table}
Table~\ref{tab:model-features} shows the features used for the parameter model. This includes characteristics for both query plan as well as the inputs to the query and is motivated by our observation that the optimal executor count depends on both of these aspects (also see Figure~\ref{fig:opt-execs}). We only use features that are available at compile-time and optimization-time of the query since (1) we want to predict the optimal executor count \textit{before} running the query and (2) we need to use the \textit{same} features for scoring the model as we used for training it. Thus, we do not include any runtime statistics as features for the parameter model. We evaluate feature importances to the model in Section~\ref{subsec:feature-importance}.

With our parametric PPM approach, we construct a single training data point for each query in our training dataset, regardless of how many different configurations for which the query run time is available. Thus, if we are training over all 103 queries of TPC-DS, our training dataset will have 103 data points. During model scoring time, the model is scored only once per query regardless of the number of candidate target configurations; the predicted times for the target configurations are determined by evaluating the PPM functions (Equations~\ref{eqn:pcc-pl} and~\ref{eqn:pcc-al}) which are generally much faster than model scoring times except those for simple, linear models. Section~\ref{subsec:overheads} discusses the overheads.

In contrast, a non-parametric approach would include run times for every configuration of each query as a separate data point in the training dataset. So, for the above example, the training dataset would have 103$\times c_{tr}$ data points where $c_{tr}$ is the number of training configurations for each query. If there are $c_{tt}$ candidate configurations for each test query, then it would score the model $c_{tt}$ times as opposed to only once with the parametric approach. Thus, our parametric PPM approach reduces training datasets, and subsequently random forest model sizes, model training and scoring times compared to a non-parametric approach.

%% file: system_integration.tex
\begin{figure*}[t]
             \centering
             \includegraphics[width=0.85\textwidth] {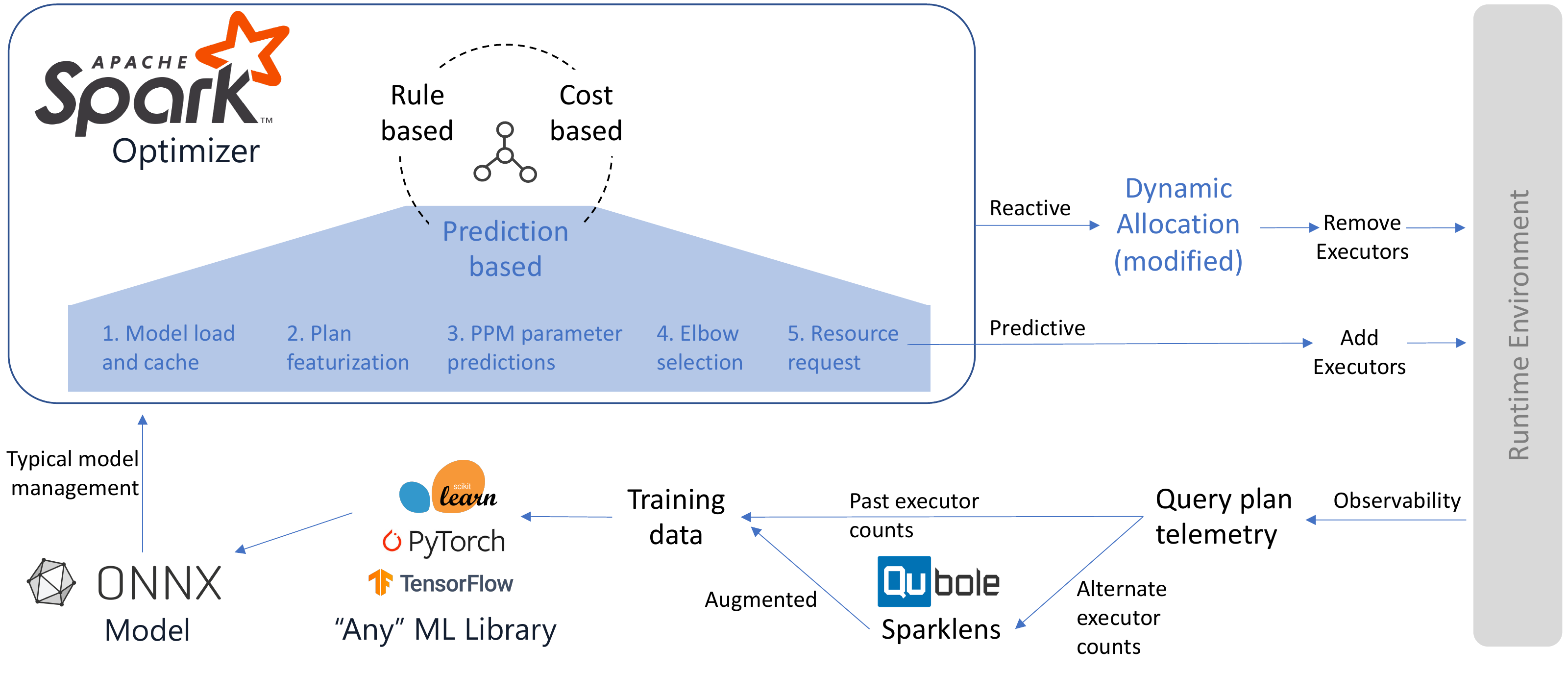}
             \caption{The \system{} system design illustrating the prediction-based resource optimization in Spark.}
             \label{fig:sys-design}
\end{figure*}

We now describe how we integrate predictive price-perf optimization with the Spark query engine~\cite{apache-spark}.
Traditionally, the Spark query optimizer performs rule-based and cost-based optimizations, and provides extensions for custom rules and cost models.
We augment the Spark optimizer to support \textit{predictive optimizations}, i.e., using ML-models to make the optimization decisions.
These models could be trained offline using any of popular ML libraries such as Scikit-learn, PyTorch, TensorFlow, etc., and we score them efficiently and in-process during the course of query optimization. We believe this a major shift for query optimization in Spark and while we focus on predictive executor counts in this paper, our approach can be leveraged for many other predictive optimizations in the future.
Figure~\ref{fig:sys-design} shows our augmented Spark optimizer with support for predictive query optimizations.
Below we walk through the different components in this architecture.

\subsection{Training Data}
We collect a rich set of data from past query runs to generate training data for \system{}.
The training features for \system include query characteristics, input dataset information, and runtime statistics as shown in Table~\ref{tab:model-features}.
To collect this information, we use Peregrine~\cite{peregrine} and SparkCruise~\cite{peregrine-spark} to log detailed plans with annotations such as input dataset information, and runtime metrics at the end of every query. 
The collected data is transformed into a tabular representation of the query workload.
The table contains one row per query. 

Given that past telemetry contains runtime metrics for a given executor count, with which the query actually ran, we need more data with different executor counts in order to train the PPM.
We achieve this by augmenting the past executions with simulated runs for other executor counts using Qubole Sparklens~\cite{sparklens}.
Sparklens simulates the Spark scheduler to provide expected runtimes with different executor counts.
These simulation points provide additional training data for the same query.
Another alternative is to re-run the queries with different executor counts.
This method is more expensive and might not be possible for production workloads.

\subsection{Model Training}
Once we have the training data that has been augmented for different executor counts, we train the parameter model described in Section~\ref{sec:ml-model}.
\system{} allows using any of the popular ML libraries, like Scikit-learn, PyTorch, TensorFlow, etc., as illustrated in Figure~\ref{fig:sys-design}. 
Although, we used Scikit-learn in this paper, this flexibility is useful in improving the models over time for different workload characteristics by trying out different libraries.
We describe the training environment and overheads in Section~\ref{subsec:overheads}.

\subsection{Model Format}
Currently, the \system pipeline uses Scikit-learn Python library to train models. 
However, the Spark optimizer code runs inside Java Virtual Machine (JVM) making it difficult to use Scikit-learn models for inference inside the optimizer. 
To solve this interoperability problem, we convert the models into ONNX format~\cite{onnx}. 
Scikit-learn, like many machine learning libraries, supports converting models into ONNX format. 
ONNX model runtime provides Java bindings and can be used inside Spark optimizer.
We can also replace the training library with TensorFlow, PyTorch, etc. as long as they also export to the ONNX model format.
Additionally, ONNX model runtime has multiple optimizations to improve the inference time. 
\system requires fast inference times as it runs inside the query optimizer and any delay will affect the end-to-end query completion time.

\begin{figure}[ht]
\centering
\includegraphics[width=0.95\columnwidth]{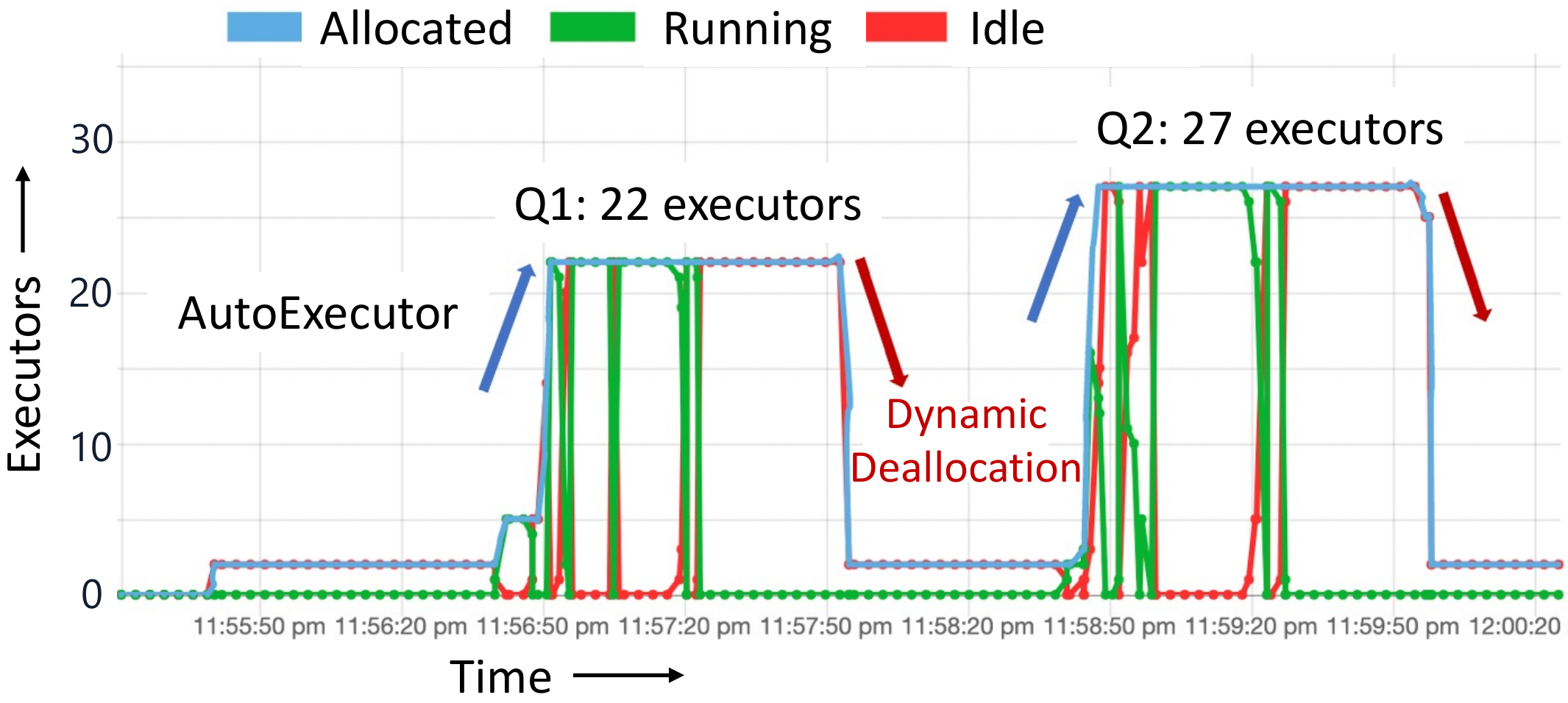}
\caption{Combining predictive and reactive resource allocation for individual queries in an interactive Spark application.}
\label{fig:sys-action}
\end{figure}

\subsection{Model Scoring}
Once the ONNX model is produced, we can leverage typical model management libraries and infrastructure, such as Azure Machine Learning or MLflow.
Thereafter, \system introduces prediction-based optimizations in the Spark query optimizer and applies a series of five steps, as illustrated in Figure~\ref{fig:sys-design}, to request the desired number of resources during optimization phase before a query is run.
We implemented the prediction-based optimizations using the Spark extensions feature~\cite{spark-ext} and it does the following.
First, we load the ONNX model from the corresponding model register, e.g., the AML model registry. 
In contrast to traditional model scoring, we load the model into the optimizer process for low-latency scoring.
We also cache the models once loaded inside the optimizer to not load them repeatedly since the inference step is in the live query path.
Then, we featurize the optimized query plan and its input datasets into a feature vector
and feed them to the parameter model to get the predicted parameters for the PPM.
With the predicted PPM parameters, \system rule gets the execution time predictions for different number of executor counts.
The default executor selection strategy automatically selects the executor count right before the performance flattens in the price-perf trade-off to achieve the fastest query performance with minimal number of executors. 
This strategy can be updated depending on the price-performance tradeoffs of the user.

\subsection{Predictive Resource Allocation}
After using the model to get the desired executor count, \system automatically requests for resources before the query is run and releases the resources after the query has finished. 
We use the executor allocation API in Spark to request additional number of executors from the cluster manager. 
Note that the allocation request is not binding and the cluster manager might allocate fewer than the requested count depending on available resources.

\subsection{Combining Predictive and Reactive Approaches}
Finally, we need to combine the predictive resource allocation with the reactive approach, namely the dynamic allocation.
Specifically, we use dynamic allocation as a mechanism to release excess resources after they are no longer needed by the query, i.e., to de-allocate resources. 
The reasoning behind this is long-running analytics queries typically require more resources upfront for more scale-out to process larger volumes of data before they get filtered or aggregated~\cite{token-shaper}. 
\system uses a modified version of dynamic allocation strategy from Spark. 
We disable dynamic allocation for scaling up since we can predict the resources upfront, but enable dynamic allocation to remove executors after they have been idle for more than a specified time duration.

Figure~\ref{fig:sys-action} shows \system with predictive allocation and reactive deallocation of executors for two queries in an interactive Spark notebook. 
When the first query is submitted, \system predicts executor count of 22 and automatically requests the predicted executor count. 
There is a time gap between the end time of first query and the submit time of second query. 
During this time gap, dynamic (de)allocation releases the idle executors. 
For the second query, \system automatically allocates the predicted executor count of 27. 
We release the resources after both the queries have finished.

%% file: evaluation.tex
We now present an experiment evaluation of \system{}. 
Our goal is to answer several key questions, including
how good are the models, 
what is the impact of price-perf trade-off,  
how much cost savings can our predictive approach bring, 
can predictive approach cope with changes in input sizes,
what are the overheads involved,
and which features are more important than others.
Below we first described out setup before discussing each of these questions.

\subsection{Setup}
\label{subsec:eval-setup}

Our testbed consists of 103 TPC-DS queries (99 queries + variants)~\cite{spark-tpcds}, for two scale factors SF=10 and 100, running on Azure Synapse Spark pools with medium-sized nodes (8 cores and 64 GB memory per node). At most two executors can be placed on each node. We allocate 4 cores ($e_c=4$) and 28 GB memory for the driver and each executor. We vary the number of executors, $n$, from 1 to 48. All other configuration parameters are fixed for these experiments. 
Each TPC-DS query runs as a separate application and we record its time elapsed, that we refer to as $t(n)$ in this work.
We run each query several times for every $n=$ 1, 3, 8, 16, 32, 48, then take averages after discarding outliers (points lying outside $\pm1.5\times$ the inter-quartile range. The `Actual' series shown in the figures correspond to this averaged run time data. 

\begin{figure}[t]
\centering
\includegraphics[width=0.85\columnwidth]{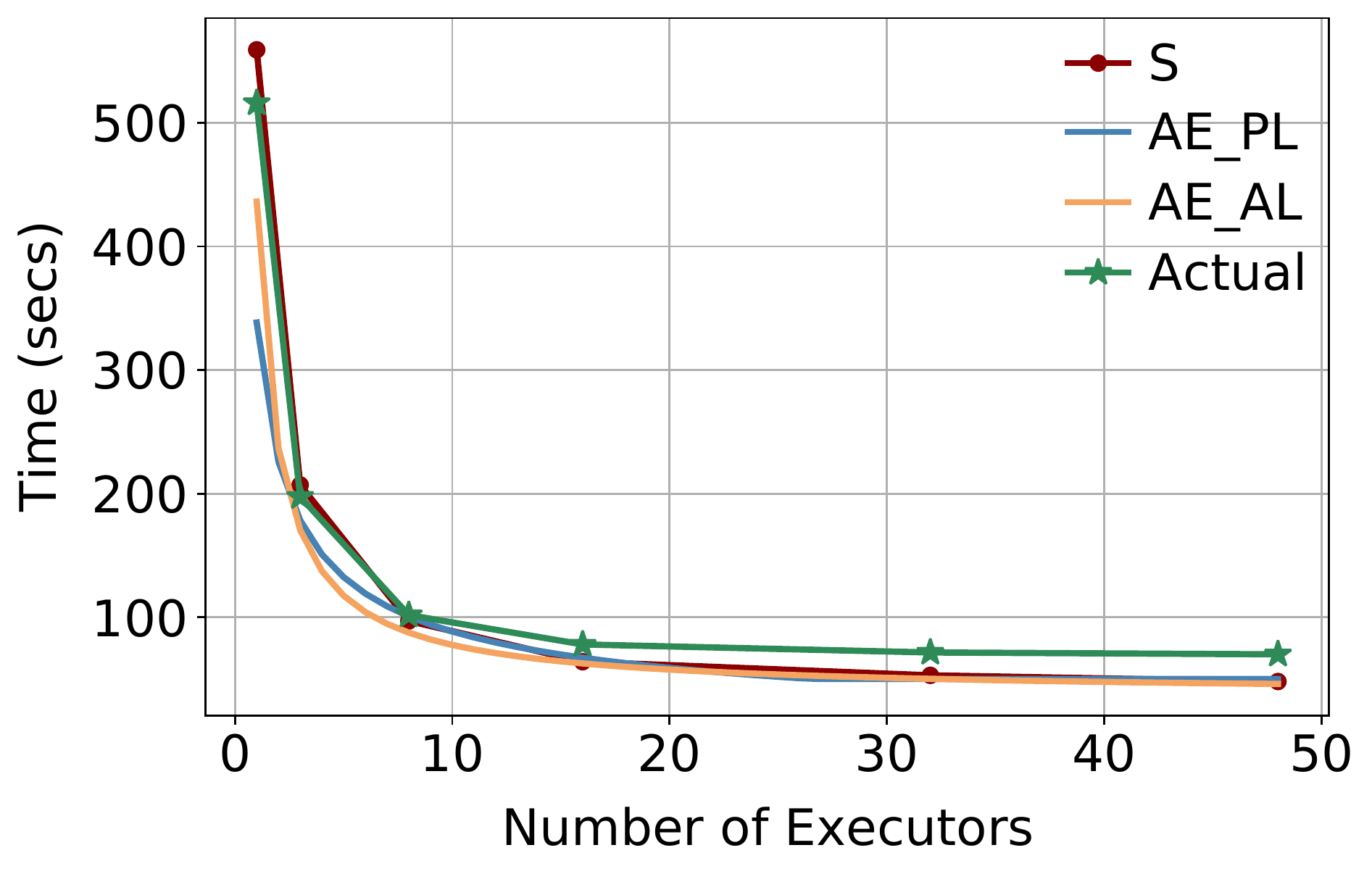}
\caption{
Comparing Sparklens estimates, PPM predictions before query execution using AE\_PL and AE\_AL models, and the actual runtimes for TPC-DS query 94, SF=100.
}
\label{fig:example-pred}
\end{figure}

\begin{figure}[ht]
\centering
\subfloat[Training Dataset]{\includegraphics[width=0.85\columnwidth]{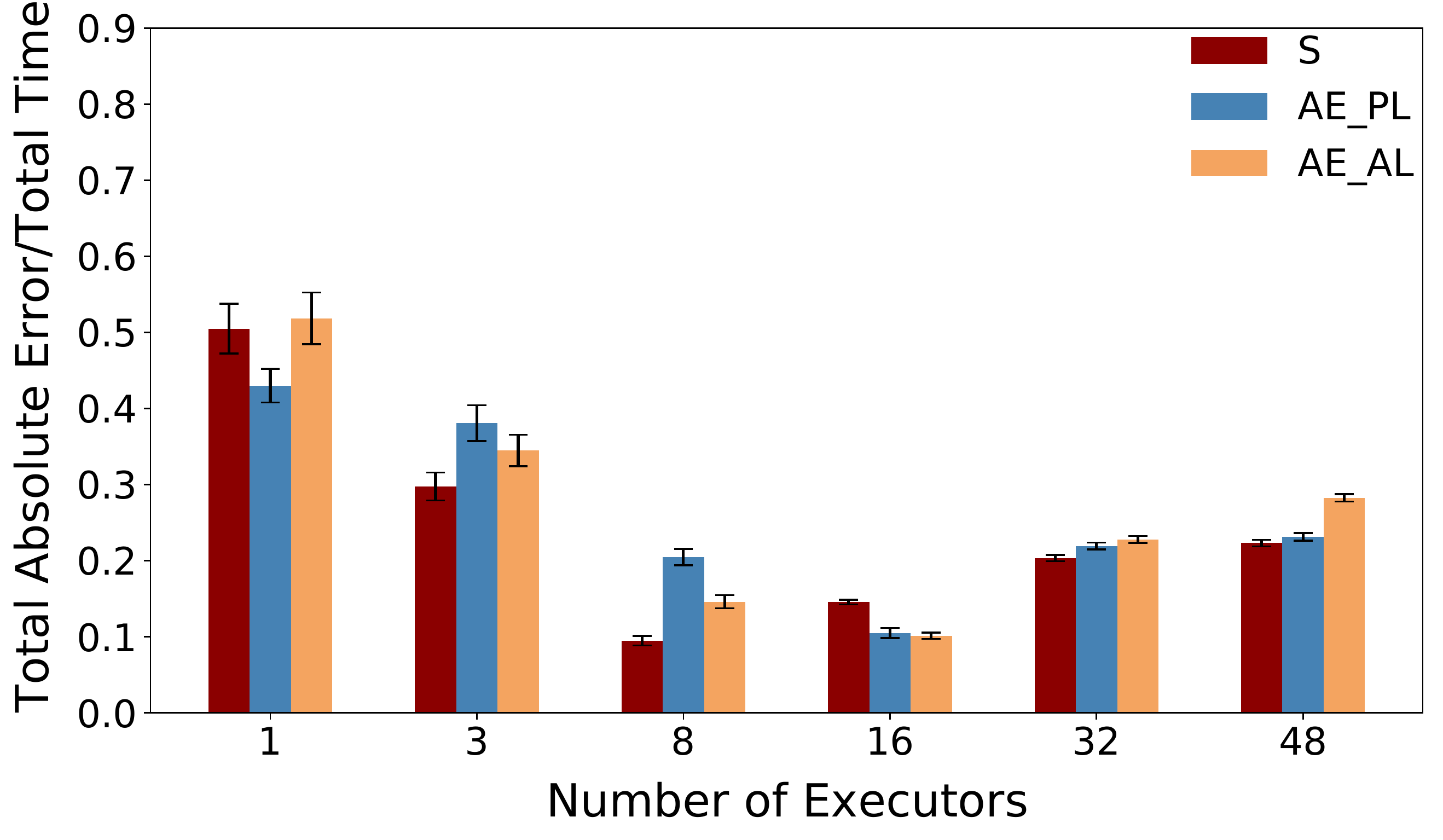}\label{fig:pred-error-tr}}\\
\subfloat[Testing Dataset]{\includegraphics[width=0.85\columnwidth]{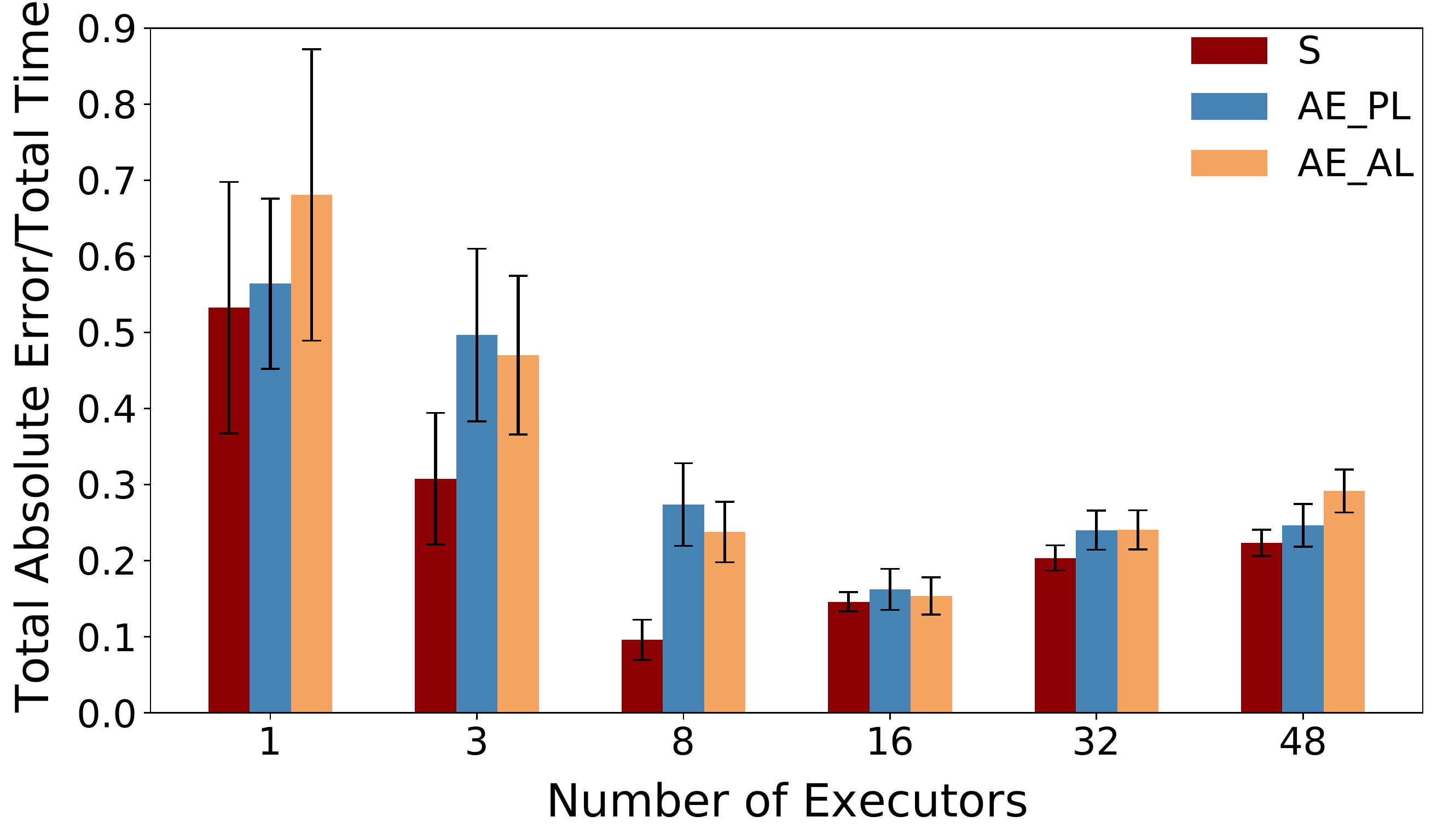}\label{fig:pred-error-tt}}
\caption{Average errors, $E(n)$, for Sparklens (S), AE\_PL, and AE\_AL aggregated over test queries from TPC-DS SF=100 (10-repeated, 5-fold cross validations). The error bars show $\pm$1 standard deviation across the $5\times10=50$ testing folds.
}
\label{fig:pred-error}
\end{figure}

We observe variation in the application run times over the different runs. 
On average, this variation (after discarding outliers) was 4.2\% (at $n=1$) to 6.9\% (at $n=48$), with a worst-case of 23.8\% (at $n=48$). Shorter run times at large $n$ tend to correlate with higher variation. Gathering ground truth data using actual runs is thus time-consuming and expensive, not only due to the need to run queries with different configurations, but also to do repeated runs to get reasonable averages. Fast and deterministic simulation tools, such as Sparklens, are thus quite useful for data augmentation to train ML models. Of course, the testing dataset only uses runtimes from actual runs.

For each query, we obtain Sparklens estimates for the application time by running the tool on the executor logs after a single run of that query with $n=16$. We show these estimates as the series 'S' in the figures.

For evaluating how well the model predictions generalize across query templates, we do a 5-fold cross validation (80:20 training:test dataset split) and repeat it 10 times. For each repeated trial, the 5 folds collectively cover all of the TPC-DS queries in the testing datasets while not including any test query in the training dataset. We evaluate two models for AutoExecutor, Power Law and Amdahl's Law, that we show as AE\_PL and AE\_AL series in the figures.

\subsection{Time Prediction}

We now evaluate how well the run times can be predicted for different values of $n$.
Figure~\ref{fig:example-pred} shows example time predictions for a query 
and compares it with the actual runtimes for that query.
Note that the training dataset for the PPM here did not include this query. We also show Sparklens estimates for comparison. However, to get the Sparklens estimates, we need to execute the query once, in this case with $n=16$. We fix $e_c=4$ for all configurations considered for this example. Although the model predictions as well as the Sparklens estimates differ from the actual run times at small executor counts, e.g., at $n=1$, the PPMs become similar at higher executor counts. Although there is a small difference in the actual execution times, the shapes of the curves are similar. 

To determine the overall prediction accuracy for the test dataset, we compute the following error metric. Let $t_q(n)$ and $\hat{t}_q(n)$ denote the actual and predicted run times for query $q$ with $n$ executors. 
\begin{equation}
E(n) = \frac{\sum_q{|\hat{t}_q(n) - t_q(n)|}}{\sum_q{t_q(n)}}
\label{eqn:E_n}
\end{equation}
Thus, $E(n)$ measures the ratio of the sum of the absolute time errors to the sum of the actual run times, with the sums taken over all queries in the test dataset. An ideal predictor would have $E(n)=0$ for all $n$.

Figures~\ref{fig:pred-error-tr} and~\ref{fig:pred-error-tt} show average $E(n)$ for the training and testing datasets corresponding to each fold of the 10-repeated 5-fold cross validations. For both models, as well as for Sparklens estimates, the errors are largest for small $n$, smallest for intermediate $n$, and intermediate for large $n$. This pattern is similar for both training (fit) and testing dataset (prediction) errors. The errors thus look to be dominated more by bias rather than by variance, and the models are not over-fitted. The relatively larger errors at small $n$ are not problematic since such $n$ are rarely optimal operating points for optimal performance or balanced tradeoff between increased cost and performance loss (`elbow points', also see Section~\ref{subsec:config-selection}). 

The model fit and predict errors are much closer to Sparklens estimation errors.
This is because we augment the training data for the PPM model with the Sparklens estimates.
The training (fit) errors show that the AE\_AL fit is better than that for AE\_PL for $n<32$, but slightly worse beyond, which is in line with our observations from Figure~\ref{fig:model-compare}. However, from the testing (prediction) errors we see that the gap is reduced and AE\_PL does better at $n=1, 48$. The average absolute difference in $E(n)$ values from Sparklens was quite small: 0.079 for AE\_PL and 0.094 for AE\_AL on the testing dataset.

\subsection{Configuration Selection}
\label{subsec:config-selection}

We now evaluate how well we can select optimal configurations based on run time predictions. We consider two scenarios: (1)~cost-savings with a limited slowdown, and (2)~elbow point selection. For these experiments, we piecewise-linearly interpolate the Actual and Sparklens series to cover all $n\in[1,48]$ and thus expand the set of target configurations. 
\begin{figure}[ht]
\centering
\subfloat[Performance]{\includegraphics[width=0.85\columnwidth]{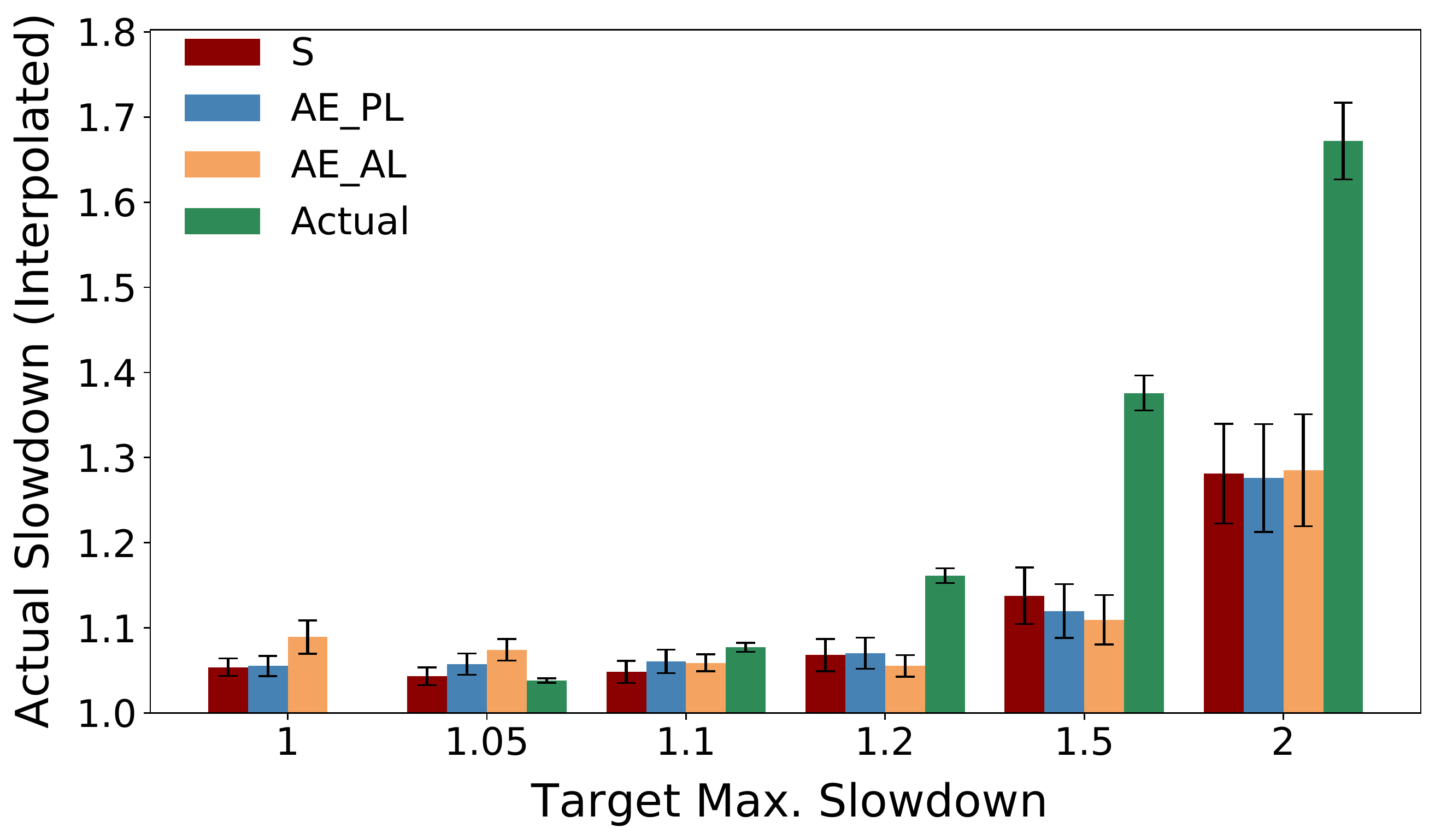}\label{fig:slowdown-H}}\\
\subfloat[Cost]{\includegraphics[width=0.85\columnwidth]{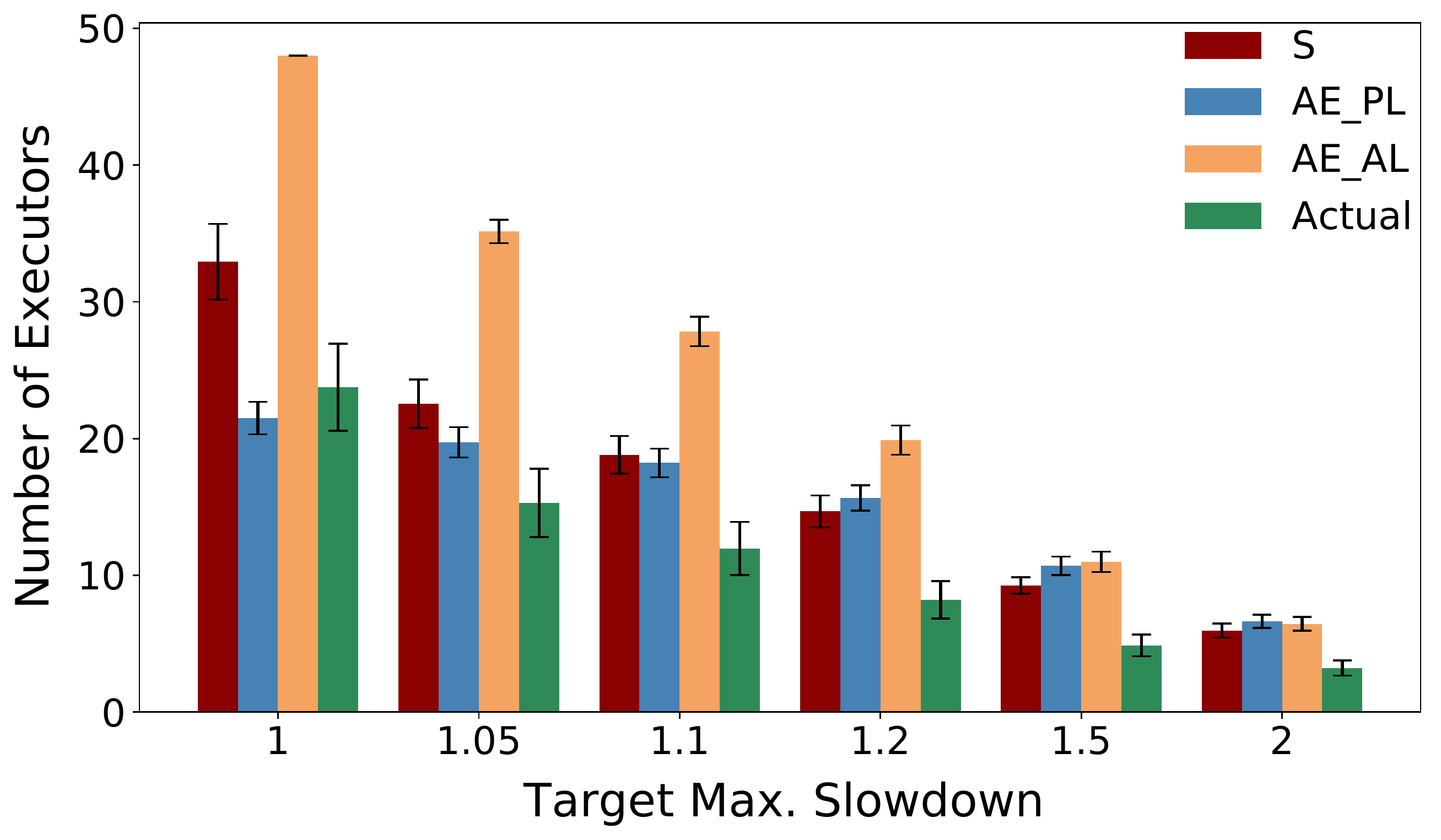}\label{fig:savings-H}}
\caption{Configuration selection impacts for Sparklens (S), AE\_PL, and AE\_AL aggregated over test queries from TPC-DS SF=100 (10-repeated, 5-fold cross validations). Slowdowns $H$ are with respect to actual run times (piecewise-linearly interpolated). The error bars show $\pm$1 standard deviation across the $5\times10=50$ testing folds.
}
\label{fig:config-sel-H}
\end{figure}
\textbf{Limited Slowdown}: The goal in this scenario is to select the \textit{smallest} $n$ such that the slowdown compared to the minimum time $t_{min}$ does not exceed a threshold $H$, i.e., $\frac{t(n)}{t_{min}}\leq H$. Figure~\ref{fig:slowdown-H} shows the slowdowns using the (interpolated) actual run times for the selected configurations for different values of $H$, while Figure~\ref{fig:savings-H} shows the corresponding $n$. We average over the queries in the test datasets for the 10-repeated 5-fold cross validations and the error bars correspond to $\pm$1 standard deviation of averages for the 10 repeats.

$H=1$ corresponds to the scenario where the smallest number of executors is used to achieve the best performance (no slowdown over $t_{min}$). We find that with the selected configurations, there is a resulting additional slowdown of 5.4\% for Sparklens, 5.5\% for AE\_PL, and 8.9\% for AE\_AL. The average values for $n$ are 24 for Actual and 32.9, 21.5, and 48 respectively for Sparklens, AE\_PL, and AE\_AL, thus reflecting a significant savings opportunity. AE\_PL improves over both Sparklens and AE\_AL and realizes a substantial portion of the savings opportunity while incurring a small additional slowdown (which is comparable to the average variation in run times). AE\_AL always select the maximum value of $n$ (= 48) due to the lack of a saturation constraint on the model unlike AE\_PL.

As discussed in Section~\ref{subsec:default-behavior}, users often run jobs with static (same settings for all jobs) and small values of $n$. Using the AE\_PL or AE\_AL models, we can automatically choose a configuration that optimizes for $H=1$ as above, that provides significant speedups over static , small values of $n$. The above selected configurations provided an average speedup of 69--70\% for static $n=3$ (12 cores) and 12.6--13.8\% for $n=8$ (32 cores). Speedups over static $n=2$ (8 cores) would be higher (expected 2.6--2.7$\times$ using interpolated values for Actuals).

For larger values of $H$, particularly $H\geq1.1$, the configurations selected by the models as also from Sparklens tend to be conservative in exploiting slowdown thresholds and saving executor counts. The average slowdowns for AE\_PL for $H=$ 1.05, 1.1, 1.2, 1.5, 2 were 1.06, 1.06, 1.07, 1.12, 1.28 for AE\_PL (with average $n=$ 19.7, 18.2, 15.7, 10.7, 6.2), but 1.04, 1.08, 1.16, 1.38, 1.67 (with average $n=$ 15.3, 12, 8.2, 4.9, 3.2) for Actual. Overall, AE\_AL tends to significantly overestimate $n$. While AE\_PL realizes only a part of the full savings potential, it has a similar impact as that of Sparklens but it is able to achieve it without executing the query as opposed to post-execution analysis by Sparklens.

\textbf{`Elbow Point' Selection}: We see in the example curves of Figures~\ref{fig:example-time-auc} and~\ref{fig:example-pred} that the PPM has two distinct regions: at low executor counts $n$, the run time $t(n)$ changes rapidly for a small change in $n$, whereas at high values of $n$, $t(n)$ changes slowly reflecting a diminishing return on investment for increasing $n$. In this configuration selection experiment, we aim to select the `elbow point' that strikes a balance between rate of decrease in $t(n)$ and rate of increase in $n$.

Note that in the PPM figures, the x- and y-axes, corresponding to $n$ and $t(n)$ respectively, are on different scales. However, we need a way to compare the two quantities in order to determine the elbow point. Our approach is to normalize both $n$ and $t(n)$ using range-scaling functions $u:n\mapsto [0,1]$ and $v:t(n)\mapsto[0,1]$, and then compute the slope at each point $u(n)$ of the normalized PPM as follows.
\begin{eqnarray}
u(n) &=& \frac{n - min(n)}{max(n) - min(n)}\\
v(t(n)) &=& \frac{t(n) - min(t(n))}{max(t(n)) - min(t(n))}\\
slope(u(n)) &=& \frac{v(t(n-1)) - v(t(n))}{u(n) - u(n-1)}
\end{eqnarray}
The elbow point $L$ is determined as the \textit{smallest} $n$ for which there is a crossover point for the slope, compared to unit slope, in the normalized PPM. That is, the smallest $n$ for which $slope(u(n))\geq 1$ and $slope(u(n+1))\leq 1$.
\begin{figure}[t]
\centering
\includegraphics[width=0.85\columnwidth]{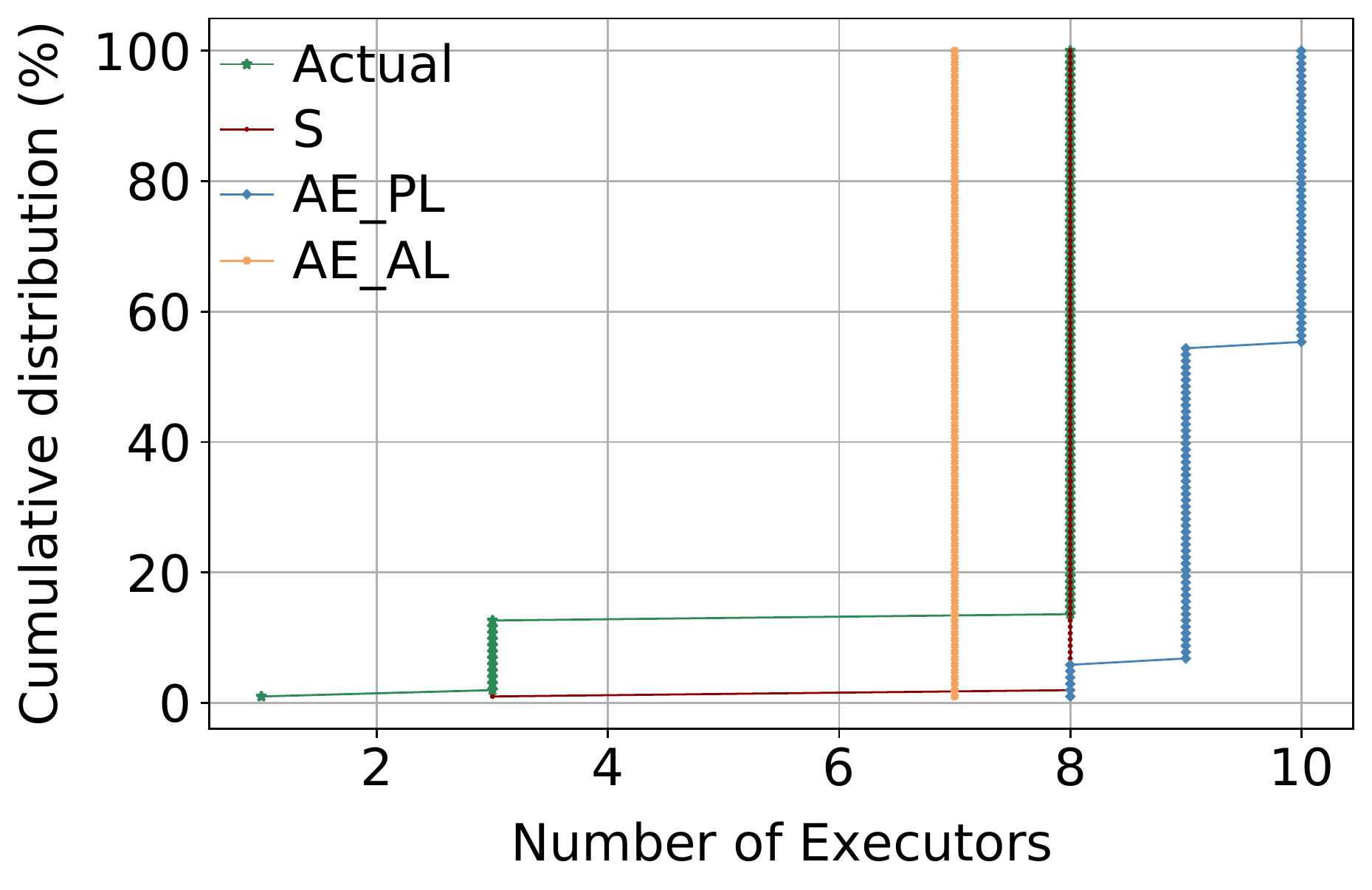}
\caption{Distribution of elbow points $L$ for all TPC-DS queries, SF=100, aggregated over the test datasets (10-repeated, 5-fold cross validations).
}
\label{fig:elbows}
\end{figure}

Figure~\ref{fig:elbows} shows the distribution of $L$ over all queries of TPC-DS, SF=100. The vast majority of queries have $L=8$; only 13 of 103 queries have $L<8$ for Actual. For Sparklens estimates, all but one query had $L=8$. For model predictions, we consider the test datasets over the 5 fold cross-validations, then take averages over the 10 repeats. Interestingly, AE\_AL always selected $L=7$ for these queries. AE\_PL selected 8, 9, or 10 for $L$. Overall, model-predicted elbow points were close to the actual values in most cases.

\subsection{Cost savings compared to static and dynamic allocation}
\label{subsec:eval-auc-savings}

\begin{figure}[t]
\centering
\includegraphics[width=0.9\columnwidth]{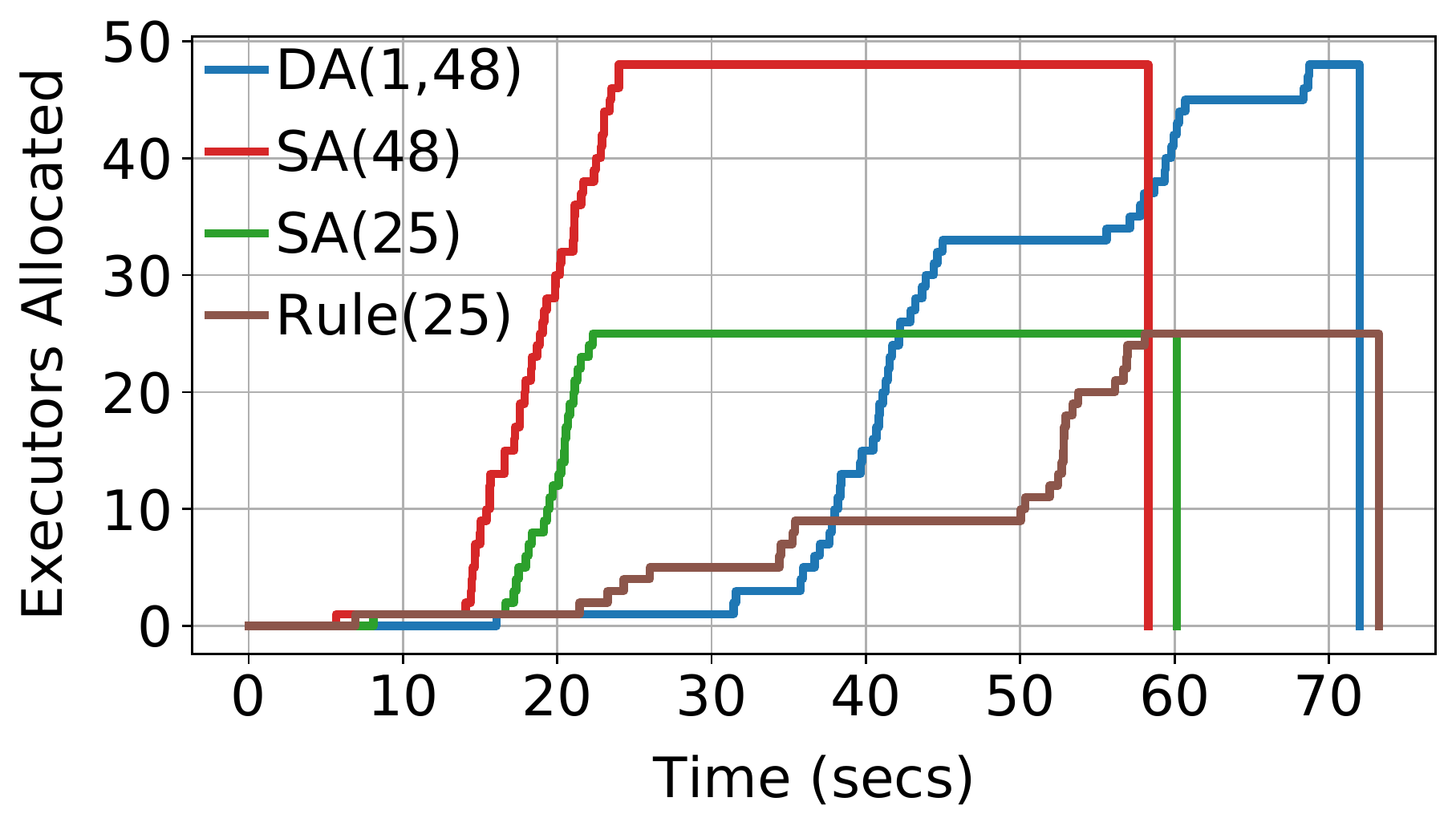}
\caption{Executor allocation skylines for TPC-DS query 94, SF=100, under Dynamic Allocation (DA) with $1\leq n \leq 48$, Static Allocation (SA) with $n=48, 25$, and SA with $n=25$ requested during optimizer Rule execution.}
\label{fig:skyines-q94}
\end{figure}
\begin{figure*}[ht]
\centering
\includegraphics[width=\textwidth]{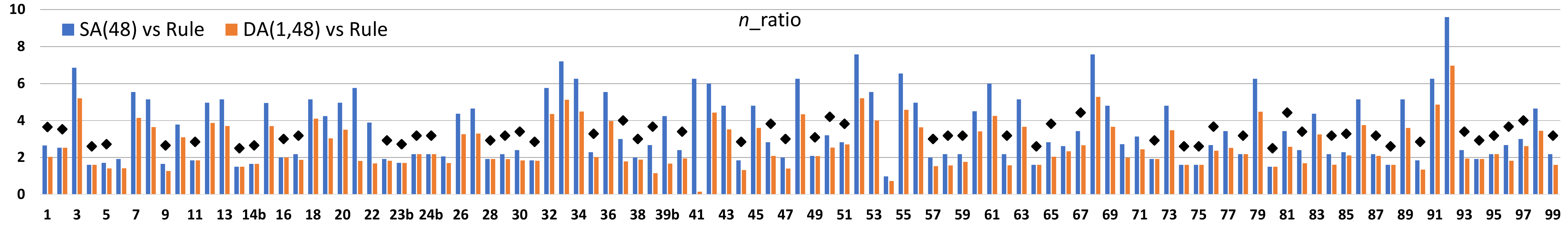}\vspace{-8pt}\\
\includegraphics[width=\textwidth]{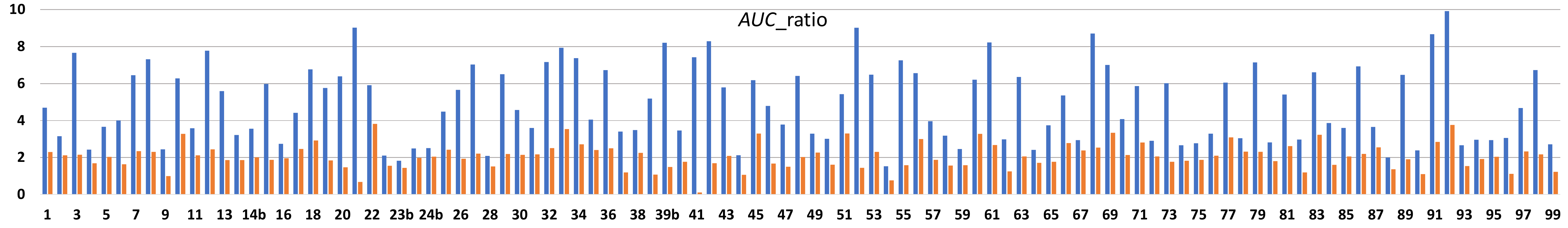}\vspace{-8pt}\\
\includegraphics[width=\textwidth]{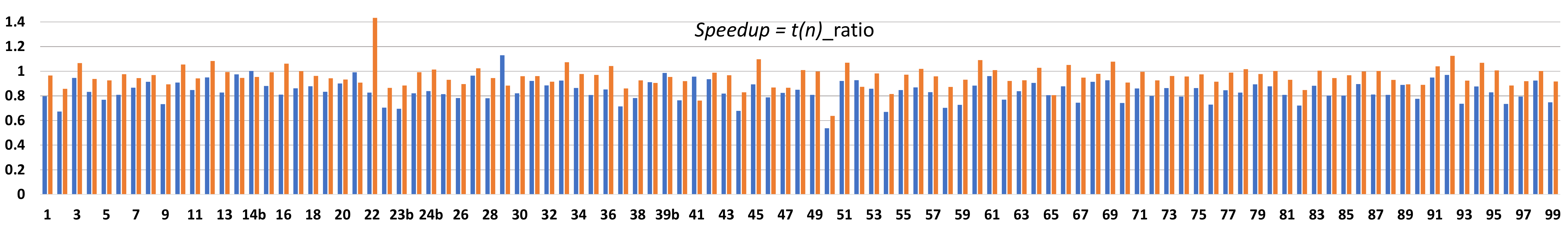}
\caption{Ratios of \texttt{DA(1,48)} and \texttt{SA(48)} to \texttt{Rule} for $n$, $AUC$, and $t(n)$ for all TPC-DS SF=100 queries where $n$ for \texttt{Rule} was selected from one of the 5-fold cross validation experiments with AE\_PL and $H=1.05$. Queries that ran long enough for all executors requested by \texttt{Rule} to be allocated by the system are marked with \ding{117}.}
\label{fig:results-detailed}
\end{figure*}
Figure~\ref{fig:skyines-q94} shows the executor allocation skylines for an example query when run using four policies: dynamic allocation (\texttt{DA}) with $n$ restricted to the range of [1, 48]; static allocation (\texttt{SA}), that is all executors requested upfront during job submission, once with $n=48$ and once with $n=25$; a request of 25 total executors made during the \system optimizer rule (\texttt{Rule}) for a run that started with $n=5$. The 25 executor count was predicted by AE\_PL for this query in one of the 5-fold cross validation experiments with optimization objective of $H=1.05$ (see Section~\ref{subsec:config-selection}). 

We note that while the run times for \texttt{SA(48)} and \texttt{SA(25)} were close, the latter substantially reduced total $n$ (48$\rightarrow$25) and $AUC$ (1904$\rightarrow$1022). Similar observations hold between \texttt{DA(1,48)} and \texttt{Rule(25)} with $AUC$ reduced from 1250 to 729. 
Both \texttt{DA} and \texttt{Rule} take more time than the \texttt{SA} policies, with the delay for \texttt{Rule} due to the lag of $\sim$27 secs from the time when the \system optimizer rule made the request and the full allocation of the requested executors by the runtime environment. 
However, \texttt{SA} policies, while faster, may only take into account query characteristics from prior runs as they need to specify $n$ even before the query is compiled.

Figure~\ref{fig:results-detailed} shows \texttt{DA(1,48)/Rule} and \texttt{SA(48)/Rule} ratios for $n$, $AUC$, and $t(n)$ over all TPC-DS SF=100 queries. \texttt{Rule} uses AE\_PL predictions for $H=1.05$ for one set of 5-fold cross validation experiments. 
The runtime environment takes $\sim$20--30 secs to gradually allocate the requested executor count. 48 of the 103 queries finished before the runtime environment was able to completely allocate all executors requested by \texttt{Rule}; we mark the rest with a \ding{117}, i.e., where the total allocated executors matched the model-predicted value. 

We see that \texttt{Rule} substantially saved $n$ and $AUC$ compared to both \texttt{SA(48)}, with average $n_{ratio}=$ 3.5 and $AUC_{ratio}=$ 4.9, and \texttt{DA(1,48)}, with average $n_{ratio}=$ 2.6 and $AUC_{ratio}=$ 2.1. The average slowdown (Speedup $< 1$) was 16\% compared to \texttt{SA(48)}, due to the lag in rule invocation and gradual allocation, but only 4\% compared to \texttt{DA(1,48)}. \system{} saved total $AUC$ by 48\% over dynamic allocation with less than 5\% performance loss, and by 73\% over static allocation.

\subsection{Change in input data size}

We now evaluate how well the models perform when the test queries operate on a different amount of data than what they had in the training datasets for the models. A simple way to test this is to train the models on one scale factor of TPC-DS and then test them on a different scale factor.

\begin{figure}[ht]
\centering
\subfloat[Testing Dataset: SF=10]{\includegraphics[width=0.85\columnwidth]{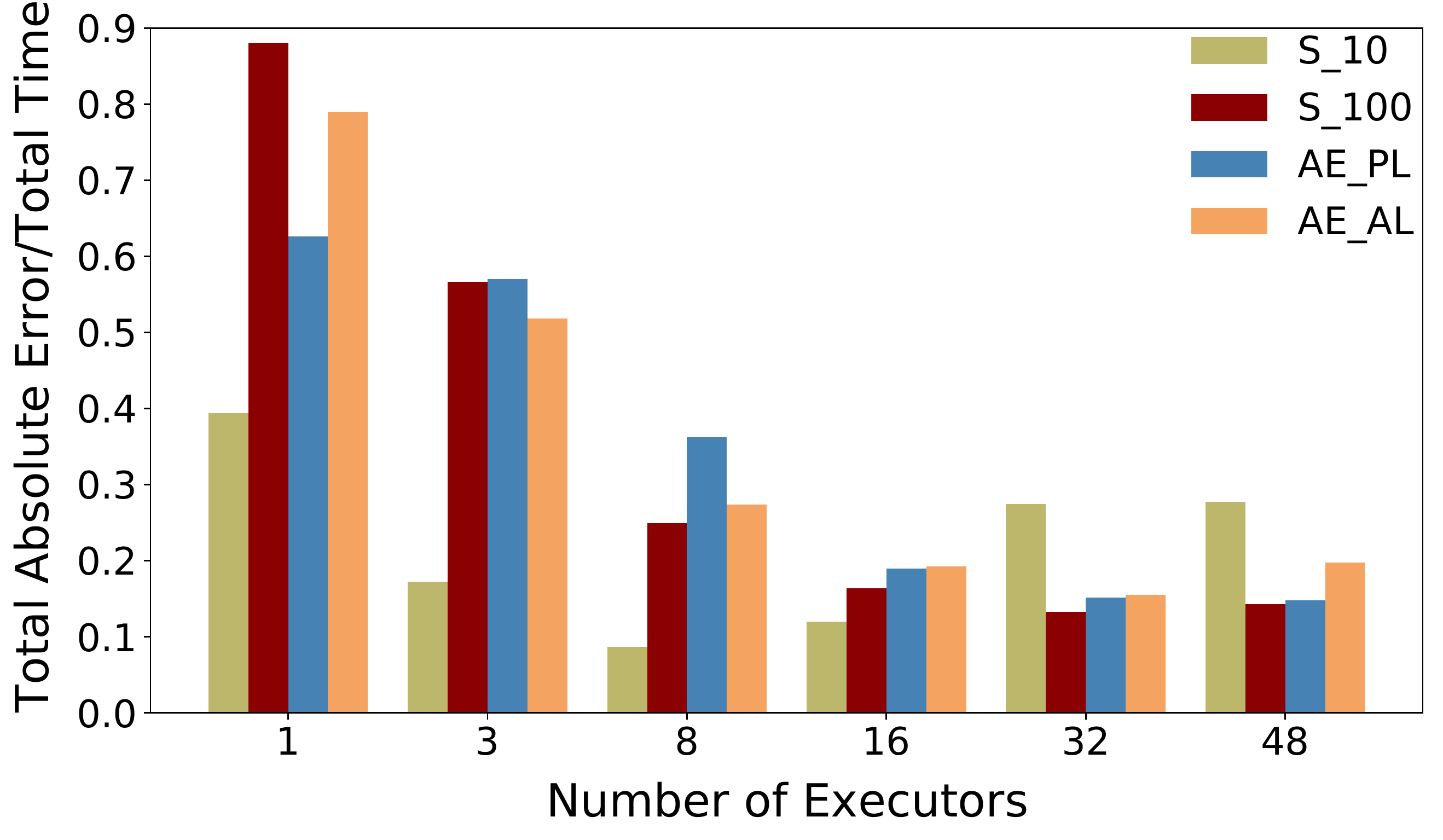}\label{fig:sf_100_10}}\\
\subfloat[Testing Dataset: SF=100]{\includegraphics[width=0.85\columnwidth]{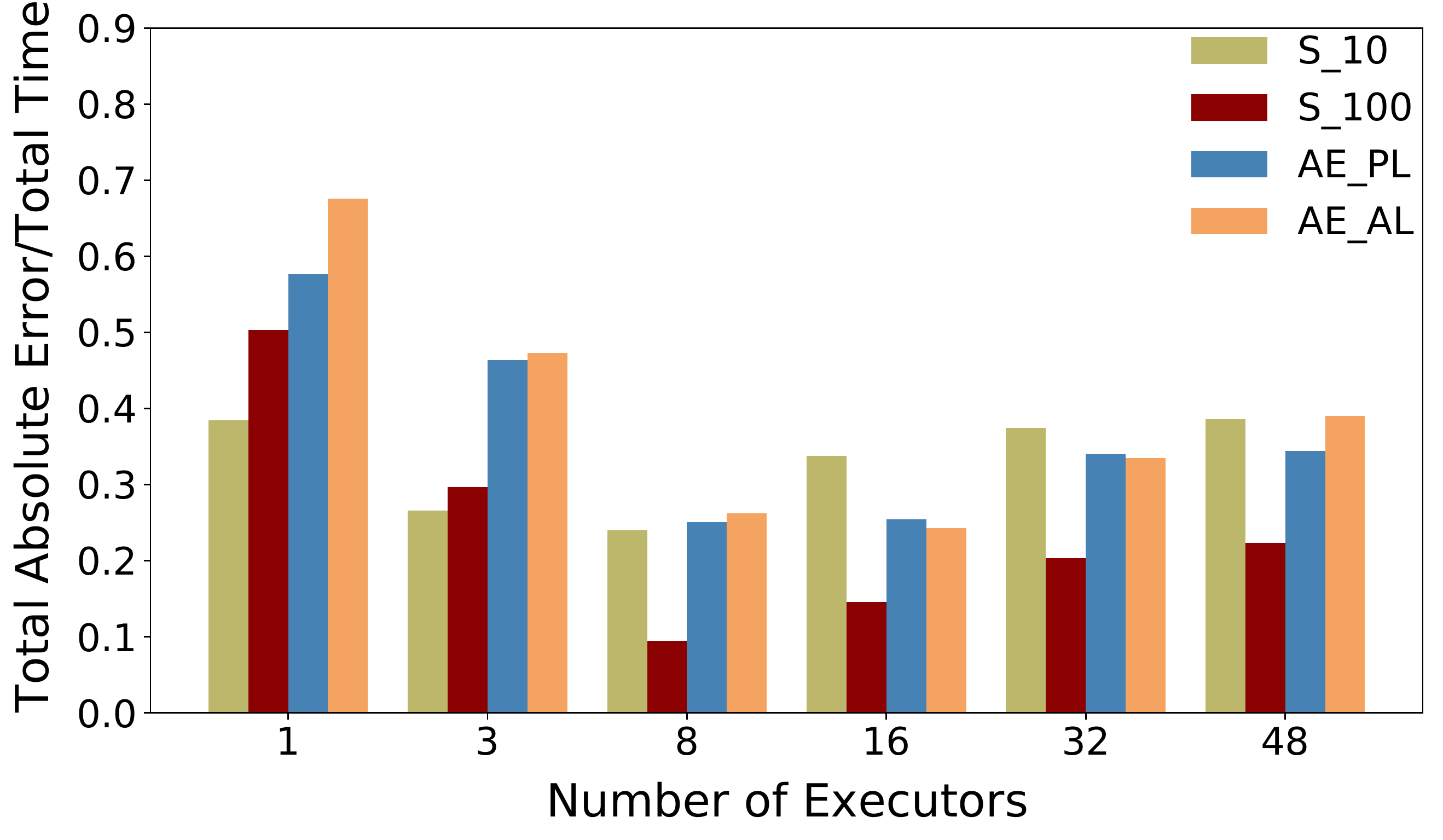}\label{fig:sf_10_100}}
\caption{Average errors, $E(n)$, for Sparklens (S), AE\_PL, and AE\_AL, aggregated over test queries from TPC-DS, SF=10 and 100. The training datasets for AE\_PL and AE\_AL are SF=100 for (a) and SF=10 for (b). S\_10, S\_100 denote the SF (= 10, 100) from which the Sparklens estimates were obtained.
}
\label{fig:pred-sf-change}
\end{figure}
Figure~\ref{fig:pred-sf-change} shows error metric $E(n)$ (see Equation~\ref{eqn:E_n}) where the testing dataset consists of all TPC-DS queries for that scale factor, which is SF=10 for Figure~\ref{fig:sf_100_10} and SF=100 for Figure~\ref{fig:sf_10_100}. In each case, the training dataset is constructed from the other scale factor, that is SF=100 and 10 respectively for the two figures. For Sparklens, we show two estimates for reference, one from running queries at SF=10 and the other from running queries at SF=10, both with $n=16$.

Overall, we see similar trends as when changing the query templates for the testing dataset (see Figure~\ref{fig:pred-error}). In particular, we see larger errors at small values of $n$ for AE\_PL as well as AE\_AL. The error ranges are also comparable for the two scenarios. Interestingly, sometimes the model predictions improve over the Sparklens estimates from the training dataset, e.g., AE\_PL vs S\_10 for testing SF=100 and $n\geq 16$. We also notice a significant difference in the average errors between S\_10 and S\_100 for both testing datasets for several $n$, e.g., S\_10 vs S\_100 for $n=8$ and testing SF=10, for $n=16$ and testing SF=100, etc. This can be understood in the light of the fact that while query behaviors change with the amount of data processed, Sparklens does not account for the change in data sizes while generating timing estimates.

\subsection{Overheads}
\label{subsec:overheads}

We now discuss various training and scoring overheads.

\textbf{Training}: The time taken to fit the parameter model on the Sparklens estimates, as described in Section~\ref{sec:ml-model}, was $\sim$0.3 msec on average for each training data point. As we have discussed, our parametric PPM approach reduces overheads by design. We used scikit-learn's default parameter settings of 100 estimators~\cite{sklearn-RF} for the Random Forest model. The pickled file size on disk when trained over all 103 TPC-DS queries (for a given scale factor) was 0.8 MB for AE\_AL and 0.9 MB for AE\_PL. The ONNX file size was slightly larger at 1 MB and 1.1 MB respectively. The average single-threaded training time for this dataset was $\sim$79 msec.

\textbf{Scoring}: The time taken to predict the PPM using the Random Forest model was on average $\sim$3.6 msec for the scikit-learn model.
Inside the query optimizer, the plan featurization time was $\sim$10.3 msec. 
The one-time cost to load and setup ONNX model for prediction was $\sim$88.1 msec and $\sim$47.1 msec respectively. 
The ONNX model inference time per query was $\sim$0.9 msec.
The \system optimizer rule is the last rule invoked once per query.

\subsection{Feature importance}
\label{subsec:feature-importance}

\begin{figure}[ht]
\centering
\includegraphics[width=0.85\columnwidth]{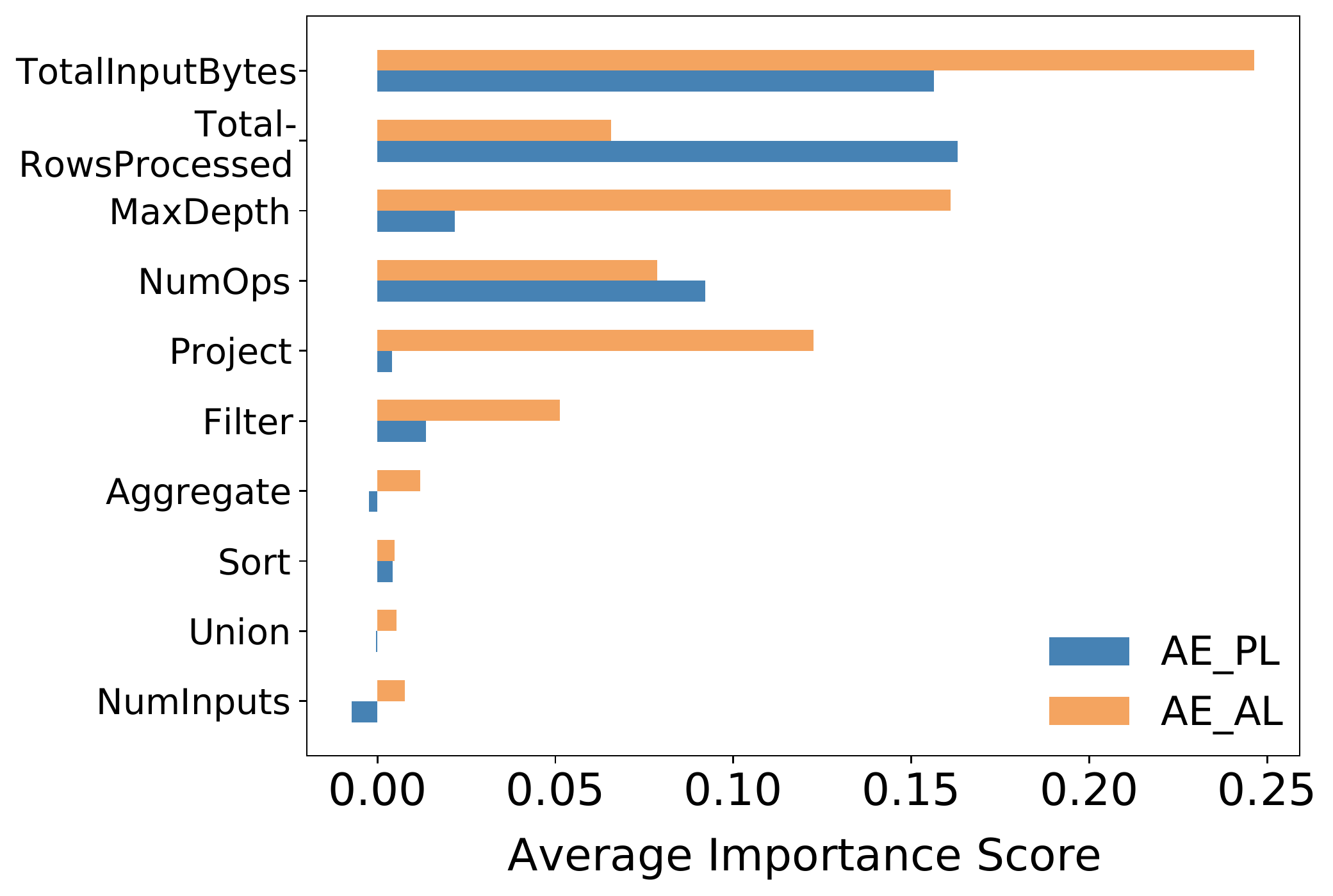}
\caption{Top 10 model features, ranked by AE\_PL + AE\_AL scores, using permutation importance for the testing datasets from TPC-DS SF=100 (10-repeated, 5-fold cross validations). For each model (fold), we repeat feature permutations 100 times. The averages are computed over $10\times5\times100=5000$ scores.}
\label{fig:feature-importance}
\end{figure}

To understand how different features contribute to model prediction accuracy, we computed feature permutation importance scores~\cite{sklearn-permutation-importance}. Figure~\ref{fig:feature-importance} lists the top 10 features in decreasing order of the sum of average importance scores for the model predictions on the testing datasets. The most important features are the estimated total input bytes and rows processed, followed by maximum depth of the query plan, number of operators, then specific operators such as Project and Filter. 
Some other operators, not shown in Figure~\ref{fig:feature-importance}, appeared much less frequently or changed less in value across queries in our dataset and got low importance scores.

To further assess the impact of these features, we performed an ablation study by constructing three reduced feature sets from the original feature set $F0$: (1) $F1$ consisting of the top six features in Figure~\ref{fig:feature-importance}, (2) $F2$ consisting of the top two features (affected by input data size), and (3) $F3 = F1 - F2$, that is, the top four plan features. Errors $E(n)$ for $F1$ were close to those for $F0$, but those for $F3$ were larger, followed by $F2$. For example, at $n=8$, $E(n)$ for $F0$, $F1$, $F2$, $F3$ were 0.27, 0.26, 0.35, 0.31 for AE\_PL and 0.24, 0.24, 0.3, 0.27 for AE\_AL. These results, as well as the feature rankings, reinforce our view that both input sizes and plan features together impact query run times and their predictions.

%% file: conclude.tex
We presented a novel approach for predictive price-performance optimization in analytical queries. 
Our system, \system, predicts a parametric model for estimating Spark SQL query run times, and picks a better resource configuration upfront during query optimization. 
\system{} can optimize for different objectives and 
is integrated with the Spark optimizer where it considers both the query characteristics and the input data sizes for predicting executor counts.
Our extensive evaluation over TPC-DS workloads show that \system{} can achieve prediction accuracies very close to Sparklens estimates, which are post-execution, and yet save $48\%$ of executor occupancy compared to dynamic allocation.